\documentclass[12pt]{article}

\begin{document}
\title{Cluster Properties in Relativistic Quantum Mechanics of 
N-Particle Systems}
\author{W. N. Polyzou \\
Department of Physics and Astronomy \\
The University of Iowa \\
Iowa City, Iowa \quad 52242 }
\maketitle
\begin{abstract}

A general technique is presented for constructing a quantum 
theory of a finite number of interacting particles satisfying
Poincar\'e invariance, cluster separability, and the spectral
condition.  Irreducible representations and Clebsch-Gordan
coefficients of the Poincar\'e group are the central elements of the
construction.  A different realization of the dynamics is obtained for
each basis of an irreducible representation of the Poincar\'e
group. Unitary operators that relate the different realizations of the
dynamis are constructed.  This technique is distinguished from other
solutions \cite{sokolov} \cite{fcwp} of this problem because it does
not depend on the kinematic subgroups of Dirac's forms \cite{dirac} of
dynamics.  Special basis choices lead to kinematic subgroups.

\end{abstract}
\bigskip
\vfill\eject
\section{Introduction}

This article illustrates a general method for constructing a relativistic
quantum theory of $N$-interacting particles.  The theory has a
dynamical unitary representation of the Poincar\'e group, satisfies
cluster separability, and has a four-momentum operator with spectrum
in the future-pointing cone.  These are the minimal elements of any
physically motivated axioms of relativistic quantum theory.
  
Relativistic quantum theory of particles falls between
non-relativistic quantum theory and local relativistic quantum field
theory.  It is interesting because it provides a mathematically
well-defined framework for realizing the symmetry of special
relativity in quantum theories.  This makes it useful for applications
to systems of a few strongly interacting particles.

The relativistic quantum theory constructed in this paper has many
properties of local relativistic quantum field theory\cite{haag}.
Both are quantum theories satisfying Poincar\'e invariance, cluster
separability, and the spectral condition.  The most significant
distinction between the two theories is that local
relativistic quantum field theory satisfies a microscopic
locality constraint, which requires an infinite number of 
degrees of freedom.

The absence of theories that are simultaneously consistent with the
axioms of local quantum field theory and applicable to realistic systems
suggests that mathematically well behaved alternatives might be well
suited to applications involving strongly interacting particles.

The essential features of quantum theory of particles are:
\begin{itemize}

\item[1.] The model Hilbert space is the finite tensor product 
of single-particle Hilbert spaces.  This defines the degrees of 
freedom of the model.

\item[2.] There is a unitary representation of the Poincar\'e group
$\hat{U}(\Lambda ,Y)$ on the model Hilbert space.  This ensures that 
the quantum probabilities are independent of inertial 
frame.  This representation necessarily contains the dynamics.

\item[3.] The four-momentum operators, which are the infinitesimal
generators of the space-time translation subgroup of $\hat{U}(\Lambda,
Y)$, have a spectrum in the future-pointing light cone.  This 
ensures the stability of the theory.

\item[4.] The operator $\hat{U}(\Lambda ,y)$ can be approximated
by a tensor product of $\hat{U}_i(\Lambda ,y)$'s on vectors describing
subsets of particles in asymptotically separated regions.   This 
justifies experiments on isolated sub-systems and provides the 
relation between few- and many-body systems. 

\item[5]  The scattering operator is unitary and Poincar\'e invariant.

\end{itemize}
 
While relativistic quantum theory of particles is useful, independent
of a relation to local quantum field theory, any local field theory
should be well approximated by a quantum theory of particles when it
is applied to reactions involving a finite number of particles.
Because the defining requirements of relativistic quantum theory of
particles are a subset of the axioms of local relativistic quantum
field theory, the consequences of these requirements on the structure
of the models of interacting particles are the same in both theories.

The Poincar\'e symmetry makes the problem of constructing a dynamical
theory difficult.  Poincar\'e covariance of the dynamics involves
non-linear constraints.  The requirement that these constraints are
preserved when the system is separated into isolated subsystems
introduces additional non-linear constraints.  These difficulties were
recognized by Dirac \cite{dirac} and have been pointed out in a recent
text by Weinberg\cite{weinberg}.

The essential role of unitary representations of the Poincar\'e group
in relativistic quantum theory was first emphasized by Wigner
\cite{wigner} in 1939.  Most applications to finite systems of
interacting particles cite Dirac's \cite{dirac} 1949 paper, which
identified the essential difficulty and introduced kinematic subgroups
associated with different ``forms of dynamics''.  These subgroups,
which reduce the number of constraints on the interactions, have
played a role in all subsequent theoretical development.
  
The problem of constructing interacting unitary representations of the
Poincar\'e group was first solved for the two-particle system by Bakamjian
and Thomas \cite{bakamjian} in 1953.  A three-particle solution satisfying
$S$-matrix clustering was given by Coester in 1965 \cite{coestera}.
The first complete solution of the problem for $N$ particles was given
by Sokolov in 1977 \cite{sokolov}.  A general solution in all of
Dirac's forms of dynamics appears in \cite{fcwp}\cite{klink}.

Relativistic quantum theory of particles is a practical framework for
applications to few-hadron
\cite{harry1}\cite{chung2}\cite{brad3}\cite{walter2}
\cite{fuda1}\cite{fuda2}\cite{salme3}\cite{klink3} and few-quark
systems
\cite{chung1}\cite{chung3}\cite{brad2}\cite{salme1}\cite{klink2}.  All
of these application are formulated in one of Dirac's forms of
dynamics; they are limited to systems where cluster
properties can be trivially realized.

The construction in this paper is directly motivated by Wigner's 1939
paper and makes essential use of irreducible representations of the
Poincar\'e group.  It generalizes the two-body construction of
\cite{polyzou} and leads to a relativistic N-body dynamics satisfying
cluster properties and the spectral condition.  Groups of unitary
transformations that preserve the $S$-matrix and cluster properties
are constructed.  In the general construction all of the Poincar\'e
generators may be interaction dependent.  The kinematic subgroup
symmetries can be implemented by imposing additional constraints on
the general construction.

The resulting dynamics has interactions in between three and ten of
the Poincar\'e generators.  Unitary operators that preserve the
S-matrix and cluster properties redistribute the interactions in ways
that may be advantageous for different applications.  These unitary 
operators are elements of a $C^*$ algebra of asymptotic constants, which 
is relevant for identifying physically equivalent theories.
 
This paper is organized as follows.  Section two contains a brief
account of Wigner's formulation of relativistic quantum mechanics,
which is central to the construction in this paper.  Sections three to
six summarize the group theory that is needed to construct the required
representations.  These sections discuss inhomogeneous $SL(2,C)\,
(ISL(2,C))$, which is the covering group of the Poincar\'e group,
irreducible representations of $ISL(2,C)$, and Clebsch-Gordan
coefficients of $ISL(2,C)$.  Section seven provides an introduction to
relativistic scattering theory, which is used in the general
construction.  This formulation of scattering theory does not assume
the existence of a kinematic subgroup.  Section eight introduces the
cluster separability condition.  Section nine introduces the $C^*$
algebra of asymptotic constants and its unitary elements, which are
called scattering equivalences.  This algebra provides a functional
calculus of non-commuting operators that is used to establish
cluster properties.  Section ten introduces the M\"obius and Zeta
function of the lattice of partitions.  These combinatoric tools,
which generalize standard Ursell cumulant expansions, are used
extensively in the construction of the $N$-body dynamics.  Section
eleven contains the general solution of the two-body problem, which is
the starting point of the recursive construction, and section twelve
contains the recursive N-body construction.  Section thirteen
constructs scattering and cluster equivalences that relate dynamical
models that utilize different bases.  Section fourteen has
conclusions.  Technical aspects of the construction are included in
the four appendices.

\section{Relativity in Quantum Mechanics}

In 1939 Wigner \cite{wigner} showed that the relativistic invariance of all
quantum probabilities
\begin{equation}
P_{\psi \phi} := \vert \langle \psi \vert \phi \rangle \vert^2 
\end{equation}
is equivalent to the existence of a unitary representation of the
Poincar\'e group.  This was refined by Bargmann in 1954 \cite{bargmann}
who observed that the dynamics could be realized by a single valued unitary
representation of the covering group, 
$ISL(2,C)$, of the Poincar\'e group.  The central problem of
relativistic quantum mechanics is to construct a unitary 
representation $\hat{U}[\Lambda ,Y]$ of $ISL(2,C)$ which implements 
the dynamics.

\section{Inhomogeneous $SL(2,C)$}

In this 
section $ISL(2,C)$ is defined and related to the Poincar\'e group. 
Elements of $ISL(2,C)$ consist of ordered pairs of 
complex $2\times 2$ matrices $(\Lambda ,Y)$, where $\Lambda$ has 
determinant 1 and $Y$ is Hermitian.  The group product is
\begin{equation}
(\Lambda_2,Y_2)(\Lambda_1,Y_1) = (\Lambda_2 \Lambda_1 , \Lambda_2 Y_1 
\Lambda_2^{\dagger} + Y_2)
\label{eq:BA}.
\end{equation}

The relation to four-dimensional Poincar\'e transformations follows 
by representing four vectors $x^{\mu}$ by $2 \times 2$ Hermitian 
matrices $X$:
\begin{equation}
X:= x^{\mu} \sigma_{\mu} \qquad 
x^{\mu} = {1 \over 2} \mbox{Tr} ( X\sigma_{\mu})
\label{eq:BB}
\end{equation}
where $\sigma_0$ is the identity and $\sigma_i$ are the Pauli matrices.
In this matrix representation $ISL(2,C)$ transformations 
are affine transformations of the form
\begin{equation}
X' = \Lambda X \Lambda^{\dagger} +Y\,. 
\label{eq:BC}
\end{equation}
Any Poincar\'e transformation continuously connected to the identity can 
be represented in the form (\ref{eq:BC}).

Elements of $ISL(2,C)$ can be parameterized by three components of a
rotation vector $\vec{\theta}$, three components of a rapidity vector
$\vec{\rho}$, and a space-time translation four vector $y^{\mu}$:
\begin{equation}
\Lambda ( \vec{\theta}, \vec{\rho}\,) =
e^{ -{i \over 2} (\vec{\theta} + i \vec{\rho}\, ) \cdot \vec{\sigma}}
\qquad
Y(y) := y^{\mu} \sigma_{\mu}. 
\label{eq:BI}
\end{equation}

Thus, the relativistic quantum dynamics, $\hat{U}[\Lambda ,Y]$,
satisfies:
\begin{equation}
\hat{U}^{\dagger}[\Lambda , Y]= \hat{U}^{-1} [\Lambda , Y] 
= \hat{U}[\Lambda^{-1} ,-\Lambda^{-1} Y (\Lambda^{-1})^{\dagger}]
\label{eq:BJ}
\end{equation}
and
\begin{equation}
\hat{U}[\Lambda_2 , Y_2] \hat{U}[\Lambda_1 , Y_1] = 
\hat{U}[\Lambda_2 \Lambda_1,
\Lambda_2 Y_1 \Lambda_2^{\dagger} + Y_2] . 
\label{eq:BK}
\end{equation}

\section{$ISL(2,C)$ Generators}

The infinitesimal generators of $\hat{U}[\Lambda ,Y]$ are defined.
These operators are used to identify a maximal set of
commuting self-adjoint operators.  For 
structureless particles the eigenvalues of
these commuting operators label the state of the particle.  The
spectrum of these operators is determined by the eigenvalues of the
invariant mass and spin operators, which define an irreducible subspace, 
and group theoretic considerations.  The single-particle Hilbert 
space is the space of square integrable functions of these
eigenvalues. 

The ten parameters $y^{\mu},\vec{\theta}, \vec{\rho}$ have the
property that if any nine of them are set to zero, the group becomes a
one-parameter unitary group with respect to the remaining parameter.
These unitary one-parameter groups necessarily have the form
$\hat{U}(\lambda )=e^{-i\lambda \hat{G}}$ for a self-adjoint operator
$\hat{G}$ \cite{reed} .  Thus a unitary
representation $\hat{U}[\Lambda ,Y]$ of $ISL(2,C)$ can be
parameterized as:

\begin{equation}
\hat{U}\left [\Lambda ( \vec{\theta}, \vec{\rho}\,), I \right]
= e^{-i ( \vec{\theta} \cdot \hat{\vec{J}} + \vec{\rho} \cdot \hat{\vec{K}})}
\label{eq:CA}
\end{equation}
\begin{equation}
\hat{U}\left[ I, Y(y) \right] = 
e^{i (\vec{y} \cdot \hat{\vec{P}}- y^0 \hat{H})}
\label{eq:CB}
\end{equation}
with self-adjoint generators $\hat{H}, \hat{\vec{P}}, \hat{\vec{J}}$ 
and $\hat{\vec{K}}$.

The commutation relations of the generators follow from the 
group representation property (\ref{eq:BK}) and the definition
$(\ref{eq:CA})(\ref{eq:CB})$ of the generators\cite{keister}.
The commutation relations are consistent with $\hat{P}^{\mu} := (\hat{H}, 
\hat{\vec{P}})$ transforming as a 
four-vector operator
\begin{equation}
\hat{U}[\Lambda ,0] \hat{P}^{\mu} \hat{U}^{\dagger}[\Lambda ,0] = 
\hat{P}^{\nu}\Lambda_{\nu}{}^{\mu}  
\label{eq:CH}
\end{equation}
and
\begin{equation}
\hat{J}^{\mu \nu} :=
\left(
\begin{array}{cccc} 
0 & \hat{K}^1 & \hat{K}^2 & \hat{K}^3 \\
-\hat{K}^1 & 0 & \hat{J}^3 & -\hat{J}^2 \\ 
-\hat{K}^2 & -\hat{J}^3 & 0 & \hat{J}^1 \\
-\hat{K}^3 & \hat{J}^2 & -\hat{J}^1 & 0 
\end{array}
\right) 
\label{eq:CG}
\end{equation}
transforming as a rank-two antisymmetric tensor operator:
\begin{equation}
\hat{U}[\Lambda ,0] \hat{J}^{\mu \nu} \hat{U}^{\dagger}[\Lambda ,0] = 
\hat{J}^{\alpha \beta}
\Lambda_{\alpha}{}^{\mu}  \Lambda_{\beta}{}^{\nu}.  
\label{eq:CI}
\end{equation}

The Pauli-Lubanski vector $\hat{W}^{\mu}$ is a four-vector valued function of 
$\hat{P}^{\mu}$ and $\hat{J}^{\mu \nu}$:
\begin{equation}
\hat{W}^{\mu} := {1 \over 2} 
\epsilon^{\mu \alpha \beta \gamma} \hat{P}_{\alpha} \hat{J}_{\beta \gamma}   
\label{eq:CJ}
\end{equation}
satisfying 
\begin{equation}
[\hat{J}^j , \hat{W}^k]_- = i \epsilon^{jkl} \hat{W}^l
\qquad
[\hat{J}^j , \hat{W}^0]_- = 0
\label{eq:CK}
\end{equation}
\begin{equation}
[\hat{K}^j , \hat{W}^k]_- =- i \delta^{jk} \hat{W}^0 
\qquad
[\hat{K}^j, \hat{W}^0]_- = -i \hat{W}^j
\label{eq:CL}
\end{equation}
\begin{equation}
[\hat{P}^{\mu}, \hat{W}^{\nu} ]_- =0 
\label{eq:CM}
\end{equation}
\begin{equation}
[\hat{W}^{\mu} , \hat{W}^{\nu}]_- = i \epsilon^{\mu \nu \rho\eta} 
\hat{W}_{\rho}\hat{P}_{\eta} 
\qquad
\hat{W}^{\mu} \hat{P}_{\mu} = 0.
\label{eq:CN}
\end{equation}
The scalar operators 
\begin{equation}
\hat{M}^2 = - \hat{P}^{\mu} \hat{P}_{\mu} 
\label{eq:CO}
\end{equation}
and
\begin{equation}
\hat{W}^2 = \hat{W}^{\mu} \hat{W}_{\mu} 
\label{eq:CP}
\end{equation}
are the two independent invariant polynomial functions of the 
generators \cite{hammermesh} of $ISL(2,C)$.

When the spectrum of the mass operator is positive, the 
spin-squared operator is defined by 
\begin{equation}
\hat{j}^2 := {\hat{W}^2 \over \hat{M}^2}.
\label{eq:CQ}
\end{equation}

\section{Irreducible Representations of $ISL(2,C)$}

The Hilbert space for an N-particle system is the tensor
product of single particle Hilbert spaces.  Single particle Hilbert
spaces are irreducible representation spaces of $ISL(2,C)$.  The
irreducible representations are labeled by the mass and spin of a
particle.  Eigenvalues of additional commuting self-adjoint functions of 
the $ISL(2,C)$ generators are needed to specify the state of the 
particle.  Simultaneous eigenstates of the 
commuting self-adjoint operators define a basis in the irreducible
representation space.  The single particle
Hilbert space is the space of square integrable functions of the 
eigenvalues.

The irreducible representations of the $ISL(2,C)$  were
classified by Wigner \cite{wigner}\cite{joos}\cite{moussa}\cite{weinberg}.  
The displacement
$x_a^\mu-x_b^{\mu}$ between events $a$ and $b$ can be classified into
six invariant classes depending on whether this displacement is zero,
lightlike positive time, lightlike negative time, spacelike, timelike
positive time, timelike negative time.

The irreducible representations corresponding to massive particles are
the timelike positive-time representations.
These irreducible representations of $ISL(2,C)$ are labeled
by the invariant eigenvalues of the mass (\ref{eq:CO}) and spin operators
(\ref{eq:CQ}).  For a particle the mass
eigenvalue $m$ is discrete and the spin operator has the eigenvalue
$j(j+1)$ where $j$ is the spin of the particle.

The state of a structureless particle of mass $m$ and spin $j$ is
determined by specifying the eigenvalues of a maximal set of commuting
self-adjoint operators.  These operators are the invariant mass $\hat{M}$, the
spin $\hat{j}^2$ and four independent functions, $\hat{F}^i = F^i
(\hat{P}^{\mu} , \hat{J}^{\mu\nu})$, of the $ISL(2,C)$ generators.
The
operators $\hat{F}^j$ cannot be invariant.  
They are arbitrary independent 
functions of the $ISL(2,C)$ generators subject to the constraints:
\begin{equation}
\hat{F}^i =  (\hat{F}^{i})^{\dagger} \qquad [\hat{F}^i,\hat{F}^j]=0
\label{eq:DA}
\end{equation}
\begin{equation}
[\hat{F}^i,\hat{M}]=[\hat{F}^i,\hat{j}^2]=0 .
\label{eq:DB}
\end{equation}
For particles with structure, additional invariant degeneracy
operators are needed to get a maximal set of commuting operators.

The traditional choice for the operators $\hat{F}^i$ are the three
components of the linear momentum $\hat{\vec{P}}$ and the
$z$-component of the canonical \cite{keister} spin $\hat{z} \cdot
\hat{\vec{j}}_c$.  In some applications it is advantageous to use the
four velocity, the light-front components of the four momentum, or
their conjugate variables.  The helicity or light-front spin is
sometimes used instead of the canonical spin.  Any of the spin
observables could be replaced by a component of the Pauli-Lubanski
operator.  These special cases are treated in Appendix I.  Each choice
of $\hat{F}^i$ corresponds to a single particle basis.  In this paper
the operators $\hat{F}^i$ are assumed to have a spectrum independent
of the mass eigenvalue.  This condition is not very restrictive and
holds for all conventional choices.

The Hilbert space for a particle of mass $m$ and spin $j$ can be
represented as the space of square integrable functions of the
eigenvalues of the operators $\hat{F}^i$:
\begin{equation}
{\cal H}_{mj} = \left\{ \langle f  \vert \psi \rangle \vert
\int d\mu (f) 
\vert \langle f \vert \psi \rangle\vert^2 \right\} < \infty
\label{eq:DC}
\end{equation}
where $f=\lbrace f^1 \cdots f^4 \rbrace$ and $ \int
d\mu(f)$ indicates a sum over the discrete eigenvalues 
and an integral over the continuous eigenvalues of $\hat{F}^i$.  

Basis vectors have the form
\begin{equation} 
\vert f \rangle := \vert f (m,j)\rangle := \vert f^1,f^2,f^3,f^4; m,j \rangle .
\end{equation}

The normalization convention is
\begin{equation}
\langle f \vert f' \rangle = \delta [f,f'] 
\label{eq:DD}
\end{equation}
where $\delta [f,f']$ is the product of Dirac or Kronecker delta functions
in the variables $f^i$.

Irreducibility requires the transformation property:
\begin{equation}
\hat{U}[ \Lambda , Y] \vert f ;m,j \rangle =
\int  \vert f' ;m,j \rangle d\mu (f') 
{\cal D}^{m,j}_{f', f } [\Lambda ,Y]
\label{eq:DE}
\end{equation}
where  
\begin{equation}
{\cal D}^{m,j}_{f',f} [\Lambda ,Y]\delta_{m'm}\delta_{j'j}   :=
\langle f' ;m',j' \vert \hat{U}[ \Lambda , Y] \vert f ;m,j \rangle 
\label{eq:DF}
\end{equation}
is the mass $m$, spin $j$ irreducible representation of $ISL(2,C)$ in 
the basis $\hat{F}^j$. 
The ${\cal D}$-function includes $\delta$-functions that eliminate the 
integrals over the continuous spectrum in $(\ref{eq:DE})$. 
Unitarity of the group representation property require: 
\begin{equation}
{\cal D}^{m,j}_{f',f } [\Lambda ,Y] =
\left ( {\cal D}^{m,j}_{f ,f'} [\Lambda^{-1} ,
- \Lambda Y \Lambda^{\dagger} ] \right )^*
\label{eq:DG}
\end{equation}
and
\begin{equation}
\int  
{\cal D}^{m,j}_{f' ,f''} [\Lambda_2 ,Y_2] d\mu (f'')
{\cal D}^{m,j}_{f'' ,f} [\Lambda_1 ,Y_1] =
{\cal D}^{m,j}_{f' ,f } [\Lambda_2 \Lambda_1  ,
\Lambda_2 Y_1 \Lambda_2^{\dagger} + Y_2].
\label{eq:DH}
\end{equation}
The restriction on the spectrum of $\hat{F}^i$ implies that
range of values of $f$ in ${\cal D}^{m,j}_{f' ,f } [\Lambda,Y]$
is independent of $m$.  

Explicit representations for the $ISL(2,C)$ Wigner ${\cal D}$-functions 
corresponding to different $\hat{F}^i$ are given in Appendix I.
The form of the ${\cal D}$-functions is basis dependent.

Irreducible representations in a basis of simultaneous 
eigenstates of a different set of commuting self-adjoint 
functions, $\hat{G}^i$, of the generators are
related to the representations in the basis $\hat{F}^i$ by:
\begin{equation}
{\cal D}^{m,j}_{g, g' } [\Lambda ,Y] =
\int \langle g \vert f \rangle 
d \mu (f) 
{\cal D}^{m,j}_{f ,f' } [\Lambda ,Y]
d\mu (f') \langle f' \vert g' \rangle
\label{eq:DI}
\end{equation}
where 
\begin{equation}
\langle f \vert g \rangle \delta_{mm'} \delta_{jj'}  :=
\langle f;m,j  \vert g;m',j' \rangle .
\label{eq:DJ}
\end{equation}
The coefficient functions $\langle f \vert g \rangle$ can depend
parametrically on the mass or spin.  This parametric dependence on the
mass is responsible for the dynamical differences that arise with
different basis choices.

\section{Clebsch-Gordan Coefficients}

In this section Clebsch-Gordan coefficients \cite{coestera}\cite{joos}
\cite{moussa}\cite{keister}  and Racah coefficients of $ISL(2,C)$ are
defined.  These are used to expand tensor products of irreducible 
representation as linear superpositions of irreducible representations 
and to transform between irreducible bases with different degeneracy 
quantum numbers.

The tensor product of irreducible representations
of $ISL(2,C)$ is reducible.  The $ISL(2,C)$ generators for a tensor 
product of two irreducible representations are 
\begin{equation}
\hat{P}^{\mu} = \hat{P}^{\mu}_1 \otimes \hat{I}_2 + 
\hat{I}_1 \otimes \hat{P}^{\mu}_2
\label{eq:EA}
\end{equation}
\begin{equation}
\hat{J}^{\mu\nu} = \hat{J}^{\mu\nu}_1 \otimes \hat{I}_2 + \hat{I}_1 \otimes 
\hat{J}^{\mu\nu}_2 .
\label{eq:EB}
\end{equation}
These operators act on the space
\begin{equation}
{\cal H} = {\cal H}_{m_1 j_1} \otimes {\cal H}_{m_2 j_2}. 
\label{eq:EC}
\end{equation}

The operators $\hat{F}^{i} =
F^i(\hat{P}^{\mu},\hat{J}^{\mu\nu})$, $\hat{M} =
M(\hat{P}^{\mu},\hat{J}^{\mu\nu})$, and $\hat{j}^2=
j^2(\hat{P}^{\mu}, \hat{J}^{\mu \nu})$ are commuting self-adjoint
operators on ${\cal H}$. Because the tensor product is reducible, these
operators do not define a maximal set of commuting self-adjoint
operators.  There are additional $ISL(2,C)$ invariant degeneracy
operators $\hat{D}^j$ that distinguish multiple copies of the same
irreducible representation.  The degeneracy operators $\hat{D}^i$
normally include the invariant operators $\hat{M}_1$, $\hat{j}_1$, $\hat{M}_2$,
$\hat{j}_2$ of the factors of the tensor product and additional
operators, $\hat{R}_{12}$, that distinguish multiple copies of the $m,j$
representation in the tensor product of the $m_1, j_1$ and $m_2, j_2$
representations.

The operators $\hat{D}^j$ are invariant, self-adjoint functions of the
single particle generators.
The operators $\hat{M}$,
$\hat{j}^2$, $\hat{F}^1 \cdots \hat{F}^4$, $\hat{D}^1,
\cdots
\hat{D}^6$ form a maximal set of commuting self-adjoint operators on 
${\cal H}$.  Examples are 
given in the Appendix II.

The $(f,d)$ basis is the $ISL(2,C)$-irreducible basis for the tensor 
product space defined in terms of simultaneous eigenstates, 
$\vert f,d ;m,j  \rangle$  of 
\begin{equation}
\lbrace \hat{F}^i , \hat{D}^k (\hat{M}, \hat{j}^2)  \rbrace .
\label{eq:EG}
\end{equation}
It follows that 
\begin{equation}
\hat{U}_1 [\Lambda,Y]\otimes \hat{U}_2 [\Lambda ,Y] 
\vert f , d  ;m,j \rangle =
\int \vert f' , d ;m,j 
\rangle d\mu (f') {\cal D}^{m,j}_{f', f } [\Lambda ,Y]
\label{eq:EH}
\end{equation}
where ${\cal D}^{m,j}_{f', f } [\Lambda ,Y]$ is the
irreducible representation matrix for a single particle 
of mass $m$ spin $j$.  The ${\cal
D}$-function is independent of the invariant degeneracy 
parameters, $d$.

The coefficients
\begin{equation}
\langle f_1 ;m_1,j_1: f_2 ;m_2,j_2) 
\vert f , d  ;m,j \rangle 
\label{eq:EI}
\end{equation}
are Clebsch-Gordan coefficients of the Poincar\'e group in the $(f,d)$ 
basis.  They are the kernel of the unitary transformation that
relate tensor products of $ISL(2,C)$ irreducible representations to
direct integrals of irreducible representations.  The $ISL(2,C)$
Clebsch-Gordan coefficients have similar properties to $SU(2)$
Clebsch-Gordan coefficients:
\[
\int {\cal D}^{m_1, j_1}_{f_1, f_1' } [\Lambda ,Y]
{\cal D}^{m_2,j_2}_{f_2, f_2' } [\Lambda ,Y] d\mu(f_1' f_2') 
\times
\]
\[
\langle f_1' ;m_1,j_1: f_2' ;m_2,j_2 
\vert f , d ; m,j \rangle =
\]
\begin{equation}
\int \langle f_1 ;m_1,j_1: f_2 ;m_2,j_2 
\vert f' , d ;m,j \rangle d\mu(f') 
{\cal D}^{m,j}_{f', f } [\Lambda ,Y] .
\label{eq:EJ}
\end{equation}
The new feature is that the irreducible representations are labeled by two 
Casimir operators and the mass operator has a 
continuous spectrum.

It is sometimes useful to replace the mass operator 
$\hat{M}$ of the tensor product 
of two irreducible representations by the invariant relative 
momentum $\hat{q}^2$, which has 
absolutely continuous spectrum, $[0,\infty )$:
\[
\hat{q}^2 = q^2(\hat{M}^2,\hat{M}_1^2,\hat{M}_2^2) :=
\]
\begin{equation}
{\hat{M}^4 + \hat{M}_1^4 +\hat{M}_2^4 - 2 \hat{M}_1^2\hat{M}_2^2 
- 2 \hat{M}^2\hat{M}_1^2 - 2 \hat{M}^2\hat{M}_2^2
\over 4\hat{M}^2}.
\label{eq:EF}
\end{equation}

The Clebsch-Gordan coefficients have different forms in different
bases.  If $(f,d)  \to (g,k)$ then the
Clebsch-Gordan coefficients in the $(f,d)$ basis are related to the
Clebsch-Gordan coefficients in the $(g,k)$ basis by
\[
\langle g_1 ;m_1,j_1: g_2 ;m_2,j_2 
\vert g , k ;m,j \rangle =
\]
\[
\int \langle g_1 \vert f_1' \rangle 
\langle g_2 \vert f_2' \rangle d\mu(f'_1)d\mu(f'_2)  
\times
\]
\begin{equation}
\langle f_1' ;m_1,j_1: f_2' ;m_2,j_2 
\vert f' , d'  ;m,j \rangle  d\mu (f' , d') 
\langle f', d' \vert g,k \rangle 
\label{eq:EK}
\end{equation}
where 
\begin{equation}
\langle g_i \vert f_i' \rangle
\delta_{j_ij_i'} \delta_{m_i m_i'}  := \langle g_i ;m_i, j_i \vert 
f_i';m_i', j_i'  \rangle 
\label{eq:EL}
\end{equation}
and 
\begin{equation}
\delta_{jj'} \delta (m-m') \langle f ,d \vert g',k' \rangle := 
\langle f ,d ;m,j \vert g',k' ;m',j' \rangle .
\label{eq:EM}
\end{equation}

The Hilbert space for a system of $N$-particles is the N-fold tensor 
product of single particle Hilbert spaces:
\begin{equation}
{\cal H} = {\cal H}_{m_1 j_1} \otimes \cdots \otimes {\cal H}_{m_N j_N}. 
\label{eq:EN}
\end{equation}

The non-interacting representation of $ISL(2,C)$ on ${\cal H}$ is 
defined by
\begin{equation}
\hat{U}_0 [\Lambda ,Y] := \hat{U}_1 [\Lambda ,Y] \otimes \cdots 
\otimes \hat{U}_N [\Lambda, Y]
\label{eq:EO}
\end{equation}
where the $0$ subscript is used to denote the non-interacting system.
It follows that 
\[
\hat{U}_0 [\Lambda, Y] \vert f_1 ;m_1, j_1 \cdots f_N ;m_N, j_N
\rangle = 
\]
\begin{equation} 
\int 
\vert f_1' ;m_1, j_1: \cdots f_N' ;m_N, j_N
\rangle d\mu (f_1' \cdots f_N')   
\prod_{i=1}^N {\cal D}^{m_i,j_i}_{f_i' ,f_i } [\Lambda ,Y].
\label{eq:EP}
\end{equation}

As in the case of $SU(2)$, the tensor product of $N$ irreducible
representation spaces can be decomposed into a direct integral of
irreducible representation spaces using successive pairwise coupling.
The invariant degeneracy operators depend on the order of the
coupling.  It is also possible to use a simultaneous coupling scheme
based on Mackey's \cite{mackey} theory of induced
representations \cite{klink} which leads to a symmetric coupling.

Successive pairwise coupling is illustrated for the three-particle system:
\[
\vert f, d_{((12)3)} ;m,j \rangle =
\]
\[
\int \vert f_1 ;m_1, j_1 : f_2 ;m_2, j_2 : f_3 ;m_3, j_3 \rangle 
d\mu (f_1) d\mu (f_2) \times
\]
\[
\langle f_1 ;m_1,j_1 : f_2 ;m_2,j_2  \vert f_{12}, d_{12} (m_{12},j_{12} )
\rangle 
d\mu (f_{12}) d\mu (f_3) d\mu (m_{12},j_{12})\times 
\]
\begin{equation}
\langle f_{12} ;m_{12},j_{12}: f_3 ;m_3, j_3\vert f, d_{12,3} ;m,j\rangle 
\label{eq:EQ}
\end{equation}
where the invariant degeneracy parameters are
\begin{equation}
d_{12,3} = \lbrace d_{12} , m_{12} , j_{12} , m_3, j_3,  
r_{12,3} \rbrace 
\label{eq:ER}
\end{equation}
with 
\begin{equation}
d_{12}= \lbrace m_1, j_1, m_2, j_2, r_{12} \rbrace .
\label{eq:ES}
\end{equation}

Changing the ordering of the coupling from $((12)3)$ to $((23)1)$ changes 
the degeneracy parameters from $\lbrace r_{12},j_{12},m_{12},r_{12,3} 
\rbrace$ to $\lbrace r_{23},j_{23},m_{23},r_{23,1} \rbrace$,
leaving the operators $\hat{M},\hat{j}^2$ and $\hat{F}^i$ unchanged.
The overlap coefficients have the general form 
\[
\langle  f, d_{ab,c} (m,j) \vert  f', d_{ef,g}' (m',j') \rangle =
\]
\begin{equation}
\delta [f, f'] \delta_{jj'} \delta (m-m') R^{m,j}_{d_{a,bc},d'_{e,fg}}.
\label{eq:ET}
\end{equation}
The invariant quantities $R^{m,j}_{d_{a,bc},d'_{e,fg}}$ are
Racah coefficients for $ISL(2,C)$.  They are the kernel of the
unitary transformation that changes the choice of degeneracy labels in
subspaces corresponding the same mass, spin, and vector labels $f$.
They are independent of $f$.

The Racah coefficients are important for performing computations because,
as in the case of rotations, some operators have a simple form
when the couplings are done in a specific order.  Since many of the operators
are defined in specific representations, the Racah coefficients 
are needed for the evaluation of the abstract operator expressions.

The term Racah coefficient is used to indicate any change of
irreducible basis with matrix elements of the form (\ref{eq:ET}).
Examples of Racah coefficients in representative bases are given in
Appendix II.

\section{Relativistic Scattering Theory} 
 
Relativistic scattering theory is formulated in this section.  A 
kinematic subgroup is not assumed.  The two-Hilbert
space formulation \cite{fcwp}\cite{coestera}\cite{chandler} is used to
treat multichannel scattering theory.  The notation of this section
follows \cite{fcwp}.  Conditions on the interactions that are
sufficient for a sensible relativistic scattering theory are
discussed.  Relativistic two-Hilbert space wave operators are
essential elements of the general construction.

In this section the dynamical representation $\hat{U}[\Lambda ,Y]$
of $ISL(2,C)$ is assumed to be given.  The construction of 
$\hat{U} [\Lambda ,Y]$ is the main topic of the remainder of this paper.

The first step in formulating relativistic scattering theory is to determine
the bound states of $\hat{U}[\Lambda, Y]$;  subsystem bound states are
needed to formulate the asymptotic conditions in multi-channel scattering.

Bound states are associated with point eigenvalues of the mass and
spin.  For each bound-state channel $\alpha_b$ there is an irreducible
subspace of ${\cal H}$.  Vectors in the bound state subspace can be
expressed as linear superpositions of simultaneous eigenstates of
$\hat{M},\hat{j}^2,\hat{F}^i$:
\begin{equation}
\vert \phi_{\alpha_b} \rangle 
= \int \vert f ;m_{\alpha} ,j_{\alpha} \rangle d\mu (f)  
\langle f \vert \chi \rangle 
\label{eq:FA}
\end{equation}
where in this expression $\hat{F}^i= F^i(\hat{P}^{\mu} , \hat{J}^{\mu \nu})$ 
are functions of the generators of $\hat{U}[\Lambda,Y]$.

The channel eigenstate $\vert f ;m_{\alpha_b} ,j_{\alpha_b} \rangle$ can be 
considered as a mapping, $\hat{\Phi}_{\alpha_b}$, from the channel Hilbert 
space ${\cal H}_{\alpha_b}$: 
\begin{equation}
{\cal H}_{\alpha_b} = \lbrace \langle f \vert \chi_\alpha \rangle  \vert
\int \vert \langle f 
\vert \chi_\alpha \rangle \vert^2 d \mu (f) < \infty \rbrace
\label{eq:FB}
\end{equation}
to the invariant bound-state subspace of  
the Hilbert space ${\cal H}$:
\begin{equation}
\hat{\Phi}_{\alpha_b} \vert \chi_\alpha \rangle := 
\vert \phi_{\alpha_b} \rangle 
= \int \vert f ; m_{\alpha} ,j_{\alpha} \rangle d\mu (f)  
\langle f \vert \chi_\alpha \rangle .
\label{eq:FC}
\end{equation}
For each bound channel $\alpha_b$ there is a channel injection operator
$\hat{\Phi}_{\alpha_b}$ and a channel Hilbert space ${\cal H}_{\alpha_b}$.
Since the bound channel spaces are irreducible representation spaces
with respect to $\hat{U}[\Lambda ,Y]$, the channel eigenstates
transform irreducibly 
\[
\hat{U}[\Lambda , Y] \vert f ;m_{\alpha_b} ,j_{\alpha_b} \rangle =
\]
\begin{equation}
\int  \vert f' ; m_{\alpha_b} ,j_{\alpha_b} \rangle d\mu (f')
{\cal D}^{m_{\alpha_b}, j_{\alpha_b}}_{f',f} [\Lambda ,Y].
\label{eq:FD}
\end{equation}

Equation (\ref{eq:FD}) can be expressed in terms of the channel 
injection operator as 
\begin{equation}
\hat{U}[\Lambda ,Y] \hat{\Phi}_{\alpha_b } = \hat{\Phi}_{\alpha_b} 
\hat{U}_{\alpha_b} [\Lambda ,Y]. 
\label{eq:FE}
\end{equation}

Scattering states are solutions of the time-dependent Schr\"odinger
equation that look like mutually non-interacting bound or elementary
subsystems in the asymptotic past or future.  To formulate
the asymptotic condition let $a$ denote a partition of $N$ particles 
into $n_a$ disjoint non-empty clusters.  Denote the $i$-th cluster 
by $a_i$ and the number of particles in the $i$-th cluster by $n_{a_i}$.  

For any partition $a$, the $N$-particle Hilbert space can be
factored into a tensor product of subsystem Hilbert spaces
${\cal H}_{a_i}$:
\begin{equation}
{\cal H} = \otimes_{i=1}^{n_a} {\cal H}_{a_i}
\label{eq:FF}
\end{equation}
\begin{equation}
{\cal H}_{a_i} = \otimes_{l \in a_i} {\cal H}_{m_l j_l}.
\label{eq:FG}
\end{equation} 

A partition $a$ has a scattering channel $\alpha$ 
if the subsystem dynamics 
\begin{equation}
\hat{U}_{a_i}[\Lambda, Y]: {\cal H}_{a_i} \to {\cal H}_{a_i}
\label{eq:FH}
\end{equation}
associated with each cluster of $a$ is either a one particle
cluster or has a bound state.

For each bound subsystem channel, $\alpha_i$, 
there is an injection operator, an asymptotic Hilbert space:
\begin{equation}
\hat{\Phi}_{\alpha_i}:{\cal H}_{\alpha_i} \to {\cal H}_{a_i}
\label{eq:FI}
\end{equation}
and an irreducible asymptotic representation $\hat{U}_{\alpha_i}[\Lambda,Y]$ 
of $ISL(2,C)$ on ${\cal H}_{\alpha_i}$ satisfying:
\begin{equation}
\hat{U}_{a_i} [\Lambda , Y] \hat{\Phi}_{\alpha_i} =
\hat{\Phi}_{\alpha_i}\hat{U}_{\alpha_i} [\Lambda , Y],
\label{eq:FJ}.
\end{equation}
These relations hold trivially for the one particle clusters. 
The asymptotic Hilbert space for the scattering channel $\alpha$ is defined 
as the tensor product of the bound channel subspaces for the
subsystems:
\begin{equation}
{\cal H}_{\alpha} = \otimes_{i=1}^{n_a} {\cal H}_{\alpha_i} .
\label{eq:FK}
\end{equation}
The channel injection operator
\begin{equation} 
\hat{\Phi}_{\alpha} :{\cal H}_{\alpha} \to {\cal H} 
\label{eq:FL}
\end{equation}
is defined by
\begin{equation}
\hat{\Phi}_{\alpha}  := \otimes_{i=1}^{n_a}\hat{\Phi}_{\alpha_i} .
\label{eq:FM}
\end{equation}
It follows from (\ref{eq:FJ})  that  $\hat{\Phi}_{\alpha}$ 
satisfies the intertwining 
relation
\begin{equation}
\hat{U}_a [\Lambda, Y ] \hat{\Phi}_{\alpha} = \hat{\Phi}_{\alpha}
\hat{U}_{\alpha }[\Lambda ,Y]
\label{eq:FN}
\end{equation}
where 
\begin{equation}
\hat{U}_a [\Lambda, Y ]:= \otimes_{i=1}^{n_a}  \hat{U}_{a_i} [\Lambda, Y ]
\label{eq:FO}
\end{equation}
and 
\begin{equation}
\hat{U}_{\alpha} [\Lambda, Y ]:= \otimes_{i=1}^{n_a}  \hat{U}_{\alpha_i} 
[\Lambda, Y ].
\label{eq:FP}
\end{equation}

In this notation a scattering state is a solution 
\begin{equation}
\vert \psi^{\pm} _{\alpha} (t) \rangle = 
\hat{U}[I,T] \vert \psi^{\pm}_{\alpha}  \rangle \qquad T:= t \sigma_0 
\label{eq:FQ}
\end{equation}
of the time-dependent Schr\"odinger equation satisfying 
the asymptotic condition
\begin{equation}
\lim_{t \to \pm \infty} \Vert \vert \psi_{\alpha}^{\pm}(t)\rangle -
\hat{U}_{a}[I,T] \hat{\Phi}_{\alpha}\vert \chi_{\alpha} \rangle
\Vert
\label{eq:FR}
\end{equation}
for $\vert \chi_{\alpha}\rangle \in {\cal H}_{\alpha}$. 

Equation (\ref{eq:FN}) can be used to express the 
asymptotic condition as
\begin{equation}
\lim_{t \to \pm \infty} \Vert \vert \psi^{\pm}_{\alpha} \rangle 
- \hat{U} [I,-T] \hat{\Phi}_{\alpha} \hat{U}_{\alpha} [I,T] 
\vert \chi_{\alpha} \rangle  \Vert =0 
\label{eq:FS}
\end{equation}
which is identically satisfied by the bound-state channels.

Equation (\ref{eq:FS}) can be expressed as
\begin{equation}
\vert \psi^{\pm}_{\alpha} \rangle := 
\hat{\Omega}_{\alpha \pm} \vert \chi_{\alpha} \rangle  
\label{eq:FT}
\end{equation}
where the channel wave operators
\begin{equation}
\hat{\Omega}_{\alpha \pm}:{\cal H}_{\alpha} \to {\cal H} 
\label{eq:FU}
\end{equation}
are defined by the strong limits
\begin{equation}
\hat{\Omega}_{\alpha \pm}:= \lim_{t \to \pm \infty}  
\hat{U} (I,-T) \hat{\Phi}_{\alpha} \hat{U}_{\alpha} (I,T).
\label{eq:FV}
\end{equation}
A sufficient condition for the existence of the channel wave operators 
is the Cook condition \cite{cook}:
\begin{equation}
\int_{t_c}^{\infty} \Vert \hat{V}_{\alpha}\hat{U}_{\alpha} 
(I,\pm T) \vert \chi \rangle \Vert dt < \infty 
\label{eq:FW}
\end{equation}
where $t_c$ is any constant and 
\begin{equation}
\hat{V}_{\alpha}  :=  \hat{H}\hat{\Phi}_{\alpha}- \hat{\Phi}_{\alpha} 
\hat{H}_{\alpha}.
\label{eq:FX}
\end{equation}

The scattering operator for scattering from channel $\alpha$ to channel
$\beta$ is the mapping from ${\cal H}_{\alpha} \to {\cal H}_{\beta}$
defined by
\begin{equation}
\hat{S}_{\beta \alpha} := 
\hat{\Omega}^{\dagger}_{\beta +}\hat{\Omega}_{\alpha -} .
\label{eq:FY}
\end{equation}

This is can be expressed compactly in a two-Hilbert space notation, where 
the asymptotic Hilbert space, ${\cal H}_{\cal A}$ is the orthogonal 
direct sum of all of the channel 
spaces, including the bound state channel spaces:
\begin{equation}
{\cal H}_{\cal A} = \oplus_{\alpha} {\cal H}_{\alpha}.
\label{eq:FZ}
\end{equation}
A two-Hilbert space injection operator $\hat{\Phi}_{\cal A}$:  
\begin{equation}
\hat{\Phi}_{\cal A} : {\cal H}_{\cal A} \to {\cal H} 
\label{eq:FAA}
\end{equation}
is defined as the sum of the channel 
injection operators
\begin{equation}
\hat{\Phi}_{\cal A} = \sum_{\alpha} \hat{\Phi}_{\alpha}  
\label{eq:FAB}
\end{equation}
where it is understood that each 
$\hat{\Phi}_{\alpha}$ acts on the channel subspace ${\cal H}_{\alpha}$ 
of ${\cal H}_{\cal A}$. 

There is a natural unitary representation of $ISL(2,C)$ on 
${\cal H}_{\cal A}$ which transforms the particles or bound states 
as tensor products of irreducible representations:
\begin{equation}
\hat{U}_{\cal A} [\Lambda ,Y] = \sum_{\alpha}  \hat{U}_{\alpha} [\Lambda ,Y]
\label{eq:FAC}
\end{equation}
where $\hat{U}_{\alpha}[\Lambda ,Y]:{\cal H}_{\alpha}\to {\cal H}_{\alpha}$.

The bound state solutions and the scattering asymptotic conditions can 
be replaced by one two-Hilbert space equation:
\begin{equation}
\Omega_{\pm}(\hat{H}, \hat{\Phi}_{\cal A} , \hat{H}_{\cal A})  = 
\lim_{t \to \pm \infty} \hat{U}[I,-T] \hat{\Phi}_{\cal A} 
\hat{U}_{\cal A} [I,T]
\label{eq:FAD}
\end{equation} 
where the limit is a strong limit. The wave operators 
$\Omega_{\pm} (\hat{H},\hat{\Phi}_{\cal A},\hat{H}_{\cal A})$
are mappings from ${\cal H}_{\cal A} \to {\cal H}$.

The scattering operator $\hat{S}$ is a mapping 
from ${\cal H}_{\cal A} \to {\cal H}_{\cal A}$
defined by 
\begin{equation}
\hat{S} := \Omega_{+}^{\dagger} (\hat{H}, \hat{\Phi}_{\cal A} , 
\hat{H}_{\cal A}) \Omega_{-}(\hat{H}, \hat{\Phi}_{\cal A} , 
\hat{H}_{\cal A}).  
\label{eq:FAE}
\end{equation}

The dynamics is asymptotically complete if the two-Hilbert space wave
operators $\Omega_{\pm}(\hat{H}, \hat{\Phi}_{\cal A} , \hat{H}_{\cal
A})$, which include all bound state channels, are unitary mappings
from ${\cal H}_{\cal A}$ to ${\cal H}$.  In all that follows the
two-Hilbert space wave operators are assumed exist and to be unitary.
These properties can be proved using the same methods used in
non-relativistic scattering theory.

Fong and Sucher \cite{fong}\cite{fcwp}\cite{polyzouc}\cite{fuda}
showed that relativistic invariance of the scattering operator does
not follow from the existence of $\hat{U}[\Lambda ,Y]$.  This is
because the $ISL(2,C)$ transformations must commute with the limiting
operations that are used to construct the scattering operator.

Invariance of $\hat{S}$ is equivalent to the condition
\begin{equation}
[\hat{U}_{\cal A} [\Lambda ,Y] , \hat{S}]_- = 0 .
\label{eq:FAF}
\end{equation}

The following theorem provides a sufficient condition on 
$\hat{U}[\Lambda ,Y]$ for the $ISL(2,C)$ invariance of the $S$-matrix:

\bigskip

\noindent {\bf Theorem 1:} Let $\Omega_{\pm} (\hat{H}, \hat{\Phi}_{\cal A}, 
\hat{H}_{\cal A})$ be asymptotically 
complete two Hilbert space wave operators.  A sufficient condition for 
$\hat{S}$ to be $ISL(2,C)$ invariant is that for all $\Lambda$ and $Y$ 
\begin{equation}
\lim_{t \to \pm \infty} \left ( \hat{\Phi}_{\cal A} - 
\hat{U}^{\dagger} [\Lambda ,Y] 
\hat{\Phi}_{\cal A} \hat{U}_{\cal A}
[\Lambda ,Y] \right ) \hat{U}_{\cal A} [I,T] =0
\label{eq:FAG}
\end{equation}
and for any $Y$ of the form $Y= \vec{y} \cdot \vec{\sigma}$ 
\begin{equation}
\lim_{t \to \pm \infty} \left ( \hat{\Phi}_{\cal A} - 
\hat{U}^{\dagger} [I ,Yt] 
\hat{\Phi}_{\cal A} \hat{U}_{\cal A}
[I ,Yt] \right ) \hat{U}_{\cal A} [I,T] =0 .
\label{eq:FAH}
\end{equation}
The limits above are strong limits. They must hold for
both time directions.

Theorem 1 provides sufficient conditions on the interactions in the
generators for a sensible relativistic scattering theory. 
The proof of this theorem is given in Appendix III. 

The proof of Theorem 1 has a number of useful corollaries:

\noindent {\bf Corollary 1} If the conditions 
of Theorem 1 are satisfied, then 
\begin{equation}
\hat{U}[\Lambda ,Y] 
\Omega_{\pm} (\hat{H}, \hat{\Phi}_{\cal A} , \hat{H}_{\cal A} ) =
\Omega_{\pm} (\hat{H}, \hat{\Phi}_{\cal A} , \hat{H}_{\cal A} )
\hat{U}_{\cal A} [\Lambda ,Y].
\label{eq:FAK} 
\end{equation}

\bigskip
This intertwining property ensures the $ISL(2,C)$ invariance of $S$.

\bigskip
\noindent {\bf Corollary 2} If the conditions 
of Theorem 1 are satisfied, then 
\begin{equation}
\Omega_{\pm} (\hat{H}, \hat{\Phi}_{\cal A} , \hat{H}_{\cal A} )=
\Omega_{\pm} (\hat{P}\cdot y, \hat{\Phi}_{\cal A} , \hat{P}_{\cal A}\cdot y )
\label{eq:FAI}
\end{equation}
where $y$ is any future-pointing time-like 4-vector.  
\bigskip

This means that all future pointing time-like directions are
equivalent for the purpose of formulating the asymptotic condition.  
\bigskip

\noindent {\bf Corollary 3}  If the conditions 
of Theorem 1 are satisfied, then 
\begin{equation}
\Omega_{\pm} (\hat{H}, \hat{\Phi}_{\cal A} , \hat{H}_{\cal A} )=
\Omega_{\pm} (\hat{M} , \hat{\Phi}_{\cal A} , \hat{M}_{\cal A} )
\label{eq:FAJ}
\end{equation}.

This shows
that in applications the Hamiltonian can be replaced by the mass
operator in the wave operators.  Both representations of the
two-Hilbert space wave operators are used in the remainder of this
paper.

\bigskip

Theorem 1 and its corollaries define conditions on the interactions that 
ensure that the dynamics is consistent with naive expectations for a 
relativistic scattering theory.  In all that follows it is assumed that
the two-Hilbert wave operators exist, are complete, and the dynamical 
operators satisfy (\ref{eq:FAG}) and (\ref{eq:FAH}). 

\section{Cluster Properties}

Cluster properties provide the essential connection between the few
and many-body problem.  The cluster property requires that
few-body interactions in the few-body problem are identical to the
few-body interactions in the many-body problem.  This establishes the
justification for performing experiments on few-body systems.

The difficulty in satisfying cluster properties is that the 
interactions that appear in the $ISL(2,C)$ generators are uniquely 
determined by cluster properties up to an $N$-body interaction.
Unfortunately, the $ISL(2,C)$ commutation relations put
non-linear constraints on the $N$-body interactions which cannot be 
satisfied by setting these interactions to zero.

To formulate cluster properties let $a$ be a partition of the $N$ 
particle systems into $n_a$ disjoint clusters.  Let $\hat{U}_{a_i}[\Lambda ,Y]$
be the subsystem representation of $ISL(2,C)$ for the particles in the 
$i$-th cluster of $a$.  Define the cluster translation operator $
\hat{T}_a (Y_1 , \cdots , Y_{n_a})$ on ${\cal H}$ by
\begin{equation}
\hat{T}_a [Y_1 , \cdots , Y_{n_a}] := 
\otimes \hat{U}_{a_i}[I ,Y_i].
\label{eq:HA}
\end{equation}

The dynamical representation of the Poincar\'e group satisfies 
strong cluster properties if for all partitions $a$ and all 
$\vert \chi \rangle \in {\cal H}$
\begin{equation}
\lim_{min (y_i-y_j)^2 \to +\infty}
\Vert  
\left (\hat{U}[\Lambda ,Y] - \otimes_{i=1}^{n_a} \hat{U}_{a_i} [\Lambda ,Y]
\right ) \hat{T}_a [Y_1, \cdots , Y_{n_a} ] \vert \chi \rangle
\Vert  =0 .
\label{eq:HB}
\end{equation}

Cluster properties will hold if (a) 
\begin{equation}
\hat{U}[\Lambda ,Y] \to U_a [\Lambda ,Y] = 
\otimes_{i=1}^{n_a} \hat{U}_{a_i} [\Lambda ,Y]
\end{equation}
when the interactions involving particles in different clusters of $a$
are set to zero and (b) all of the interactions in each generator 
$\hat{G}$ satisfy: 
\begin{equation}
\lim_{min (y_i-y_j)^2 \to +\infty}
\Vert  
\left ( \hat{G} - \hat{G}_a
\right ) \hat{T}_a [Y_1, \cdots , Y_{n_a} ] \vert \chi \rangle
\Vert  =0 
\label{eq:HBX}
\end{equation}
where $\hat{G}$ and $\hat{G}_a$ are the $ISL(2,C)$ generators
associated with $\hat{U}[\Lambda ,Y]$ and $U_a [\Lambda ,Y]$
respectively.

Condition (a) is called the algebraic cluster property
\cite{fcwp}.  It puts the non-linear constraints on the interactions of
a relativistic quantum theory.  It ensures that once the interactions
between particles in different clusters are turned off the remainder
is a tensor product.  This condition is non-trivial because it must
hold for every possible clustering.

The condition (b) is related to the range of the interaction.  If
the operators satisfy algebraic cluster properties the proof of the
short range condition is similar to the non-relativistic proof
\cite{simon2} of cluster properties.   In all that follows the 
interaction terms are assumed to satisfy condition (b). 

When $\hat{U}[\Lambda,Y]$ does not satisfy algebraic cluster properties the
limit (\ref{eq:HB}) may not exist.  A typical consequence is  
that the cluster limit eliminates interactions between 
particles in the {\it same} cluster \cite{keister} .  

The cluster condition (\ref{eq:HB}) is a strong form of the cluster 
condition.  It is also possible to formulate a weaker form of the cluster 
condition that applies only to the scattering matrix \cite{fcwp}.
The stronger form is needed for the recursive construction in sections
12 and 13.

\section{Scattering Equivalences}

There is a large class of dynamical models with the same $S$-matrix.
These models are called scattering equivalent models
\cite{fccatholic}.  The freedom to transform between scattering
equivalent models with different properties is an important tool for
realizing cluster properties.  What separates scattering equivalent
models from unitary equivalent models is that scattering equivalent
models do not change the description of free particles.  They provide
a parameterization of the freedom that is created by restricting the
class of physical observables to asymptotic quantities ($t \to \pm
\infty$).

While scattering equivalences necessarily preserve cluster properties
of the $S$-matrix, they do not preserve cluster properties of the
representation $\hat{U}[\Lambda ,Y]$.  Because of this property,
scattering equivalences can be used to restore cluster properties of
the dynamics.

The key to understanding scattering equivalences is to understand the 
algebra of operators that are asymptotically zero.  A bounded
operator $\hat{Z}$ on the $N$-particle Hilbert space is 
asymptotically zero if the following strong limits vanish
\begin{equation}
\lim_{t \to \pm \infty} \hat{Z} \hat{U}_0 [I,T] \vert \psi \rangle =0; 
\label{eq:GA}
\end{equation}
\begin{equation}
\lim_{t \to \pm \infty} \hat{Z}^{\dagger}  \hat{U}_0 [I,T] \vert \psi \rangle =0; 
\label{eq:GB}
\end{equation}
for {\it both} time limits,  where
\begin{equation}
T= t\sigma_0.
\label{eq:GC}
\end{equation}
  
The subset of bounded operators that are asymptotically zero
are denoted by ${\cal Z}$.  It is straightforward to show that for
$\hat{Z}_n \in {\cal Z}$ and $\alpha$ complex that
\begin{equation}
\alpha \hat{Z}_1 + \hat{Z}_2  \in {\cal Z} 
\label{eq:GD}
\end{equation}
\begin{equation}
\hat{Z}_1\hat{Z}_2  \in {\cal Z}
\label{eq:GE}
\end{equation}
\begin{equation}
\hat{Z}_1^{\dagger} \in {\cal Z} 
\label{eq:GF}
\end{equation}
\begin{equation} 
\Vert \hat{Z}_n - \hat{Z} \Vert \to 0 \Rightarrow \hat{Z} \in {\cal Z}.
\label{eq:GG}
\end{equation}
Including the identity makes a $C^*$ algebra, which we call the 
algebra of asymptotic constants, ${\cal C}$.  

A scattering equivalence $\hat{A}$ is a unitary member of ${\cal C}$ that is 
asymptotically equal to the identity $\hat{I}$:
\begin{equation}
\lim_{t \to \pm \infty} (\hat{A} - \hat{I} )  \hat{U}_0 [I,T] \vert \psi \rangle =0; 
\label{eq:GH}
\end{equation}
\begin{equation}
\lim_{t \to \pm \infty} (\hat{A}^{\dagger} - \hat{I} )  
\hat{U}_0 [I,T] \vert \psi \rangle =0; 
\label{eq:GI}
\end{equation}

The relation of these operators to scattering is through the following 
theorems:

\bigskip
\noindent {\bf Theorem 2:} Let $\hat{A}$ be a scattering equivalence.
Let $\Omega_{\pm} (\hat{H}, \hat{\Phi}_{\cal A}, \hat{H}_{\cal A})$
be asymptotically complete two Hilbert space wave operators.
Let $\hat{H}' = \hat{A} \hat{H} \hat{A}^{\dagger}$ and $\hat{\Phi}'_{\cal A}  
= \hat{A} \hat{\Phi}_{\cal A}$.  Then 
$\Omega_{\pm} (\hat{H}', \hat{\Phi}'_{\cal A}, \hat{H}_{\cal A})$
exist, are asymptotically complete, and give the same $S$ matrix as 
$\Omega_{\pm} (\hat{H}, \hat{\Phi}_{\cal A}, \hat{H}_{\cal A})$.

\bigskip
The proof follows from the identity
\begin{equation}
\hat{A} \Omega_{\pm} (\hat{H}, \hat{\Phi}_{\cal A}, \hat{H}_{\cal A}) = 
\Omega_{\pm} (\hat{H}', \hat{\Phi}_{\cal A}', \hat{H}_{\cal A}) .
\label{eq:GJ}
\end{equation}
While the structure of the injection operator $\hat{\Phi}_{\cal A}$ 
depends on the
representation of the subsystem bound states, it must become the
identity in the scattering channel, $(\alpha = \alpha_0)$,
corresponding to $N$ free particles.  Note that $\hat{\Phi}_{\alpha_0}'= 
\hat{A} \hat{I} = 
\hat{I}+\hat{Z} \not= \hat{I}$ where $\hat{Z}$ is
asymptotically zero.  This ensures that $\hat{\Phi}'_{\cal A}$ can be
replaced by another injection operator, $\hat{\Phi}''_{\cal A}$, 
with $\hat{\Phi}''_{\alpha_0
}= \hat{I}$:
\begin{equation}
\hat{\Phi}''_{\cal A} = 
\hat{\Phi}'_{\cal A} - \delta_{\alpha \alpha_0}\hat{Z} .  
\end{equation}
It follows that 
\begin{equation}
\Omega_{\pm}(\hat{H}', \hat{\Phi}_{\cal A}', \hat{H}_{\alpha_0}) 
=
\Omega_{\pm}(\hat{H}', \hat{\Phi}_{\cal A}'', \hat{H}_{\alpha_0}) 
\label{eq:GK}
\end{equation}
where $\hat{\Phi}_{\alpha_0}'' = \hat{I}$.

Scattering equivalences are naturally constructed from pairs of wave
operators that give the same $S$-matrix. 

\bigskip

\noindent {\bf Theorem 3:} Let 
$\hat{\Omega}_{\pm}:=\Omega_{\pm} (\hat{H}, \hat{\Phi}_{\cal A} , 
\hat{H}_{\cal A})$ and 
$\hat{\Omega}'_{\pm}:=\Omega_{\pm} (\hat{H}', \hat{\Phi}'_{\cal A} , 
\hat{H}_{\cal A})$ be asymptotically complete wave operators 
that give the same scattering matrix.  Then there is a scattering
equivalence $\hat{A}$ satisfying $\hat{H}' = \hat{A} \hat{H}
\hat{A}^{\dagger}$.

\bigskip

To prove Theorem 3 note that the assumptions imply
\begin{equation}
S = \hat{\Omega}_{+}^{\dagger}\hat{\Omega}_{-} =
\hat{\Omega}_{+}^{\prime \dagger}\hat{\Omega}'_{-}.
\label{eq:GL}
\end{equation}
Asymptotic completeness implies
\begin{equation}
\hat{A} := \hat{\Omega}'_{+}\hat{\Omega}_{+}^{\dagger}  =
\hat{\Omega}'_{-}\hat{\Omega}_{-}^{\dagger}.
\label{eq:GM}
\end{equation}
This definition and the intertwining relations \cite{simon2} for 
the Hamiltonian 
give
\[
\hat{A} \hat{H} \hat{A}^{\dagger} =
\hat{\Omega}'_{+}\hat{H}_{\cal A} \hat{\Omega}_{+}^{\dagger} \hat{A}^{\dagger}=
\]
\begin{equation}
\hat{H}' \hat{\Omega}'_{+} \hat{\Omega}_{+}^{\dagger} \hat{A}^{\dagger}=
\hat{H'}.
\label{eq:GN} 
\end{equation}
Equations (\ref{eq:GM}) and (\ref{eq:GN}) imply
\begin{equation}
\Omega_{\pm}  (\hat{H}' , \hat{\Phi}'_{\cal A} , \hat{H}_{\cal A} ) =
\hat{A} \Omega_{\pm}  (\hat{H} , \hat{\Phi}_{\cal A} , \hat{H}_{\cal A} )=
\Omega_{\pm}  (\hat{H}' , \hat{A} \hat{\Phi}_{\cal A} , \hat{H}_{\cal A} ).
\label{eq:GO}
\end{equation}
The equality of the first and last terms gives the strong limit
\begin{equation}
\lim_{t \to \pm \infty}  (\hat{\Phi}'_{\cal A} - \hat{A} \hat{\Phi}_{\cal A})
\hat{U}_{\cal A} [I,T] =0.
\label{eq:GP}
\end{equation}
Unitarity of $\hat{A}$  gives
\begin{equation}
\lim_{t \to \pm \infty}  (\hat{A}^{\dagger} \hat{\Phi}'_{\cal A} - 
\hat{\Phi}_{\cal A})
\hat{U}_{\cal A} [I,T] =0, 
\label{eq:GQ}
\end{equation}
restricting to the $\alpha_0$ channel, using $\hat{\Phi}_{\alpha_0} = 
\hat{\Phi}'_{\alpha_0}=\hat{I}$ and $\hat{U}_{\alpha_0} [I,T]= \hat{U}_0 [I,T]$
gives
\begin{equation}
\lim_{t \to \pm \infty}  (\hat{A} - \hat{I}) 
\hat{U}_{0} [I,T] =0 
\label{eq:GR}
\end{equation}
and
\begin{equation}
\lim_{t \to \pm \infty}  (\hat{A}^{\dagger} - \hat{I}) 
\hat{U}_{0} [I,T] =0 
\label{eq:GS}
\end{equation}
which establishes that $\hat{A}$ is a scattering equivalence.  

This shows that if two asymptotically complete
wave operators give the same scattering matrix then 
the Hamiltonians are related by a scattering equivalence.  Since $\hat{A}$
is unitary it follows that 
\begin{equation}
\hat{U}'[\Lambda ,Y] := \hat{A} \hat{U}[\Lambda ,Y] \hat{A}^{\dagger} 
\label{eq:GT}
\end{equation}
is a scattering equivalent representation of $ISL(2,C)$.

The important property of the scattering equivalences is that they are
the unitary elements of the $C^*$ algebra of asymptotic constants.
The $C^*$ algebra can be used to construct functions of the
non-commuting scattering equivalences.  When these functions are
unitary and can be expressed expressed as uniform limits of elements
of this algebra, they are scattering equivalences.  This provides
the mechanism for constructing scattering equivalences with
specialized properties.

\section{\bf Birkhoff Lattices:}

The construction of operators satisfying cluster properties 
requires a significant amount of algebra involving 
cluster expansions of operators.  
The theory of Birkhoff lattices
\cite{birkhoff}\cite{rota}\cite{polyzoua}\cite{kowalski}\cite{fcwp}  
facilitates the required algebra.  It provides closed-form expressions 
relating different standard cluster expansions of operators. 

Let ${\cal P}$ denote the set of all possible partitions of N-particles into 
disjoint non-empty clusters.  There is a natural partial ordering on 
${\cal P}$ given by
\begin{equation}
a \supseteq b  
\label{eq:IA}
\end{equation}
if and only if every pair of particles in the same cluster of $b$ is 
in the same cluster of $a$. This means that $b$ can be obtained from $a$
by breaking up clusters.

The Zeta and M\"obius functions \cite{rota}\cite{kowalski} for this partial 
ordering are integer valued
functions on ${\cal P} \times {\cal P}$ defined by
\begin{equation}
\zeta (a \supseteq b ) =
\left \{
\begin{array}{ll}
1 & a \supseteq b \\
0 & \mbox{otherwise} 
\end{array}
\right.
\label{eq:IB}
\end{equation}
and
\begin{equation}
\mu (a \supseteq b ) = \zeta^{-1} (a \supseteq b ) =
\left \{
\begin{array}{ll}
(-)^{n_a} \prod_{i=1}^{n_a} (-)^{n_{b_i}}(n_{b_i}-1)! & a \supseteq b \\
0 & \mbox{otherwise} 
\end{array}
\right.
\label{eq:IC}
\end{equation}
where $n_{b_i}$ are the number of clusters of $b$ in 
the $i$-th cluster of $a$.
Note that both $\zeta (a \supseteq b)$ and $\mu (a \supseteq b )$
vanish unless $a \supseteq b$.

Intersections and unions, $a \cap b$ and $a \cup b$, of two partitions
$a$ and $b$ are defined as the greatest lower bound and least upper
bound with respect to this partial ordering.

It follows from the definitions that 
\begin{equation}
\zeta ((a \cap b) \supseteq c ) = \zeta (a \supseteq c )
\zeta (b \supseteq c )
\label{eq:ID}
\end{equation}
\begin{equation}
\zeta (a \supseteq (b \cup c )) = \zeta (a \supseteq b )
\zeta (a \supseteq c ).
\label{eq:IDD}
\end{equation}

The set of partitions with the operations $\cup$ and $\cap$ form a
semimodular lattice  \cite{birkhoff}, called a partition or Birkhoff 
lattice.  It provides a convenient means for 
keeping track of interactions.  Let ${\cal O}$ be an operator that is a
function of the physical $ISL(2,C)$ infinitesimal generators.
Imagine putting a parameter $\lambda_i$ in front of each interaction 
that appears in the the physical $ISL(2,C)$ generators.  
The operator ${\cal O}_a$ is defined to be the result of turning off
the interactions between particles in different clusters of $a$.  In
general the operator ${\cal O}_a$ will include the contributions of
operators in ${\cal O}_b$ for all $a \supset b$.  These can be
recursively subtracted to construct truncated contributions 
$[{\cal O}]_b$ to
${\cal O}_a$.  The truncated operators $[{\cal O}]_a$ vanish whenever 
interactions involving particles in {\it any} cluster of $a$ are
turned off.   The the M\"obius
function can be used to generate closed form expressions for the 
truncated operators in terms of the untruncated ${\cal O}_a$'s: 
\begin{equation}
[{\cal O}]_a := \sum_{b} \mu (a \supseteq b ) {\cal O}_b .
\label{eq:IE}
\end{equation}
This can be inverted using the Zeta function to get
\begin{equation}
{\cal O}_a := \sum_{b} \zeta (a \supseteq b ) [{\cal O}]_b .
\label{eq:IF}
\end{equation}
If this is applied to the case where $a$ is the 1-cluster partition,
this becomes
\begin{equation}
{\cal O} = \sum_b [{\cal O}]_b .
\label{eq:IG}
\end{equation}

While this generates the standard relations between ordinary multipoint
functions and truncated multipoint functions based on cluster expansion
methods, use of the lattice structure, and specifically the underlying
partial ordering, has advantages that are useful in the 
recursive construction described in sections 12 and 13.

\section{\bf Two-Body Problem} 

The construction of two-body models follows \cite{polyzou}.
The two-body Hilbert space is the tensor product of single particle
spaces
\begin{equation}
{\cal H} = {\cal H}_{m_1 j_1}\otimes {\cal H}_{m_2 j_2} . 
\label{eq:JA}
\end{equation}

Choose a basis $( f, d)$ and use the Clebsch-Gordan
coefficient:
\begin{equation}
\langle f_1 ;m_1, j_1: f_2 ;m_2,j_2 \vert f,d ;m,j \rangle
\label{eq:JB}
\end{equation}
to construct an irreducible free-particle basis.  
The states 
\begin{equation}
\vert f,d ;m,j \rangle
\label{eq:JC}
\end{equation}
transform as mass $m$ spin $j$ irreducible representations of $ISL(2,C)$
with respect to the non-interacting representation 
\begin{equation}
\hat{U}_0 [\Lambda ,Y] := \hat{U}_{1}  [\Lambda ,Y] \otimes 
\hat{U}_{2}  [\Lambda ,Y].
\label{eq:JD}
\end{equation}

Using the $ISL(2,C)$ transformation 
properties it is possible to construct operators $\Delta \hat{F}_0^i$ 
that change the value of $f^i$, holding the values of $f^j$, $(j\not= i)$ 
constant.  If $\hat{F}_0^i$ has a continuous 
spectrum these operators
are proportional to partial derivatives 
\begin{equation}
\Delta \hat{F}_0^j = i {\partial \over \partial f^j} 
\label{eq:JE}
\end{equation}
holding $f^k; k\not= j$ constant.  If $F_0^j$ has discrete 
eigenvalues, a suitable $\Delta \hat{F}_0^j$ can typically be expressed 
in terms of a raising or lowering operators.

The operators $\hat{M}_0$ , $\hat{j}^2_0$, $\hat{F}_{0}^i$, $\Delta 
\hat{F}_{0}^i$ 
are functions of the free particle generators.  Expression for the generators 
in terms of these operator can be constructed using the $ISL(2,C)$  
${\cal D}$-functions:
\[
\langle f,d ;m,j\vert \vec{K}_0 \vert f',d' ;m,s \rangle := 
\]
\begin{equation}
i{\partial \over \partial \vec{\rho}} {\cal D}^{m,j}_{f,f'} [ \Lambda
(\theta=0 , \rho), 0 ] \delta [d,d'] \delta (m-m') \delta_{jj'}
\label{eq:JF}
\end{equation}

\[
\langle f,d ;m,j\vert \vec{J}_0 \vert f',d' ;m,j \rangle :=
\]
\begin{equation} 
i{\partial \over \partial \vec{\theta}} {\cal D}^{m,j}_{f,f'} [
\Lambda (\theta , \rho=0), 0 ] \delta [d,d'] \delta (m-m') \delta_{jj'}
\label{eq:JG}
\end{equation}

\[
\langle f,d ;m,j\vert P_0^{\mu}  \vert f',d' ;m,j \rangle := 
\]
\begin{equation}
-i{\partial \over \partial y_{\mu} } {\cal D}^{m,j}_{f,f'} [ I, Y(y) ]
\delta [d,d'] \delta (m-m') \delta_{jj'}
\label{eq:JH}
\end{equation}
where all derivatives are computed at $0$.

The chain rule gives explicit expressions
for the $ISL(2,C)$ generators in terms of the operators 
$\hat{M}_0, \hat{j}^2_0 , \hat{F}_{0}^i , 
\Delta \hat{F}_{i0}$:
\begin{equation}
\hat{P}^{\mu}_0 = \hat{P}^{\mu} (\hat{M}_0, \hat{j}_0 , \hat{F}_{0}^i , 
\Delta \hat{F}_{0}^i )
\label{eq:JI}
\end{equation}
\begin{equation}
\hat{J}^{\mu \nu}_0 = \hat{J}^{\mu \nu} (\hat{M}_0, \hat{j}_0 , 
\hat{F}_{0}^i ,\Delta \hat{F}_{0}^i ).
\label{eq:JJ}
\end{equation}
These expressions can be inverted to express 
$\hat{M}_0, \hat{j}^2_0 , \hat{F}_{0}^i , 
\Delta \hat{F}_{i0}$ in terms of
the $ISL(2, C)$ generators:
\begin{equation}
\hat{M}_0 = M (\hat{P}^{\mu}_0 , \hat{J}^{\mu \nu}_0 )
\end{equation}
\begin{equation}
\hat{j}^2_0 = j_0 (\hat{P}^{\mu}_0 , \hat{J}^{\mu \nu}_0 )
\end{equation}
\begin{equation}
\hat{F}_{0}^i = F^j  (\hat{P}^{\mu}_0 , \hat{J}^{\mu \nu}_0 )
\end{equation}
\begin{equation}
\Delta \hat{F}_{i0} = \Delta F^j (\hat{P}^{\mu}_0 , \hat{J}^{\mu \nu}_0 ).
\label{eq:JJA}
\end{equation}
Examples of these operators for specific basis 
choices are computed in Appendix I to illustrate the general
procedure.

Since $\hat{M}^2_0$ is a Casimir operator for $ISL(2,C)$, it necessarily
commutes with $\hat{j}^2_0 , \hat{F}_{0}^i, $ and $\Delta
\hat{F}_{0}^i$.  The $ISL(2,C)$ commutation relations follow
as consequences of the commutation relations of $\hat{M}_0,
\hat{j}^2_0 , \hat{F}_{0}^i, $ and $\Delta \hat{F}_{0}^i$. 

It follows that in order to construct a dynamical representation of
$ISL(2,C)$ it is enough to replace $\hat{M}_0$ by an operator
$\hat{M}= \hat{M}_0 + \hat{V}$ which also commutes with 
$\hat{j}^2_0 , \hat{F}_{i0}, $ and $\Delta \hat{F}_{i0}$.
With this choice of interaction it follows that the operators 
\begin{equation}
\hat{P}_0^{\mu} \to \hat{P}^{\mu} = \hat{P}^{\mu} (\hat{M}, \hat{j}^2_0 , 
\hat{F}_{0}^i , 
\Delta \hat{F}_{0}^i )
\label{eq:JK}
\end{equation}
\begin{equation}
\hat{J}_0^{\mu \nu} \to \hat{J}^{\mu \nu} = 
\hat{J}^{\mu \nu} (\hat{M}, \hat{j}^2_0 , \hat{F}_{0}^i , 
\Delta \hat{F}_{0}^i )
\label{eq:JL}
\end{equation}
automatically satisfy the $ISL(2,C)$ Lie algebra.

Cluster properties are satisfied for sufficiently short-range interactions. 
For the interaction to be non-trivial it should also satisfy
\begin{equation}
[\hat{M}, \hat{M}_0] \not=0 
\label{eq:JM}
\end{equation}
and the spectral condition,  $\hat{M}_0 > \hat{V}$.   In general 
the interaction 
can be treated as a perturbation of different functions 
of $\hat{M}_0$, such as 
$\hat{M}_0^2$. In 
all cases the interactions can be put in the form $\hat{M}=
\hat{M}_0 +\hat{V}$
by defining $\hat{V}:=\hat{M}-\hat{M}_0$, independent of how 
$\hat{M}$ is constructed.   The spectral condition constrains the 
interaction.

In the free particle irreducible basis an interaction $\hat{V}$
commuting with $\hat{j}^2_0 , \hat{F}_{0}^i, $ and $\Delta \hat{F}_{0}^i$
has a kernel with the structure:
\[
\langle  f,d ;m,j \vert \hat{V} \vert f',d' ;m',j' \rangle
=
\]
\begin{equation}
\delta [f,f'] \delta_{jj'} \langle d, m \Vert \hat{V}_j \Vert d', m' 
\rangle .
\label{eq:JO}
\end{equation}

The dynamical generators are given by (\ref{eq:JK}) and (\ref{eq:JL})
with $\hat{M}=
\hat{M}_0 +\hat{V}$.
If the expression for a generator in (\ref{eq:JK}) or (\ref{eq:JL}) 
has an explicit 
mass dependence,  the corresponding operator will be interaction 
dependent.  Depending on the choice of basis $(f,d)$ between
three and ten generators will have an explicit interaction dependence.
Dirac's forms of dynamics result from specific basis choices.
A generic choice will not have a kinematic subgroup.

While it is straightforward to derive explicit expressions for the 
generators in terms of the $\hat{F}^i$'s, (see Appendix I) 
it is easier to directly 
solve for the dynamics in the free particle 
basis $\vert f,d ;m,j \rangle$.

In this basis $\hat{M}$, $\hat{F}_{0}^i$, $\hat{j}_0$ can be simultaneously 
diagonalized:
\begin{equation}
\langle  f',d' ;m',j' \vert f ;m,j \rangle = \delta [f,f']\delta_{jj'} 
\phi_m^j ( d',m' )
\label{eq:JT}
\end{equation}
where $\phi_m^j ( d',m' )$ is the solution of the mass eigenvalue 
equation 
\begin{equation}
(m-m') \phi_m^j ( d',m' ) = \sum \int dm'' dd'' 
\langle d', m' \Vert \hat{V}_j \Vert d'' m'' \rangle \phi_m^j ( d'',m'' ).
\label{eq:JU}
\end{equation}
For suitable interactions $\hat{M}$ will be self-adjoint and the
eigenstates $\vert f,d ;m,j \rangle $ will define a complete set of
simultaneous eigenstates of $\hat{M}$, $\hat{F}^i_{0}$, $\hat{j}^2_0$.
Solving equation (\ref{eq:JU}) is of comparable difficulty to solving
the time-independent non-relativistic 
Schr\"odinger equation.  It is assumed that the 
eigenstates include two-body bound states and scattering states 
satisfying incoming and outgoing wave asymptotic conditions.

Since the expressions (\ref{eq:JF}-\ref{eq:JH}) for the $ISL(2,C)$ 
generators were derived by evaluating the infinitesimal transformations 
in an irreducible basis, and $\{ \hat{M}_0, \hat{F}_{0}^i, 
\Delta \hat{F}^i,\hat{j}_0\}$ and 
$\{\hat{M},\hat{F}_{0}^i, 
\Delta \hat{F}^i, \hat{j}_0\}$ have the same commutation relations,
the action of the dynamical representation of $ISL(2,C)$ on the eigenstates 
$ \vert f ;m,j \rangle $ has the same form 
as the free dynamics on $\vert f,d;m_0,j \rangle$, with the eigenvalue 
of $\hat{M}_0$ replaced by the eigenvalue of 
$\hat{M}$.  It follows that 
\begin{equation}
\hat{U}[\Lambda , Y] \vert f ;m,j \rangle = 
\vert f' ;m,j \rangle d\mu (f') {\cal D}_{f',f}^{m,j} [\Lambda ,Y] .
\label{eq:JV}
\end{equation}
Since the states $\vert f ;j,m \rangle$ 
are complete, this defines $\hat{U}[\Lambda ,Y]$ on ${\cal H}$.
Since $m$ is the eigenvalue of a dynamical operator, all of the mass
dependent parts of ${\cal D}_{f'f}^{mj} [\Lambda ,Y]$  are interaction 
dependent.

This construction gives (1) an explicit expressions for the interaction
dependent $ISL(2,C)$ Lie algebra, (2) a solution of the $2$-body
dynamics expressed as a direct integral of irreducible representations of
$ISL(2,C)$, (3) and an explicit unitary representation of $ISL(2,C)$ on ${\cal
H}$.

This construction can be done in any irreducible basis.  Consider the
same construction in two bases $(f,d)$ and $(g,d)$ where, for simplicity,
the degeneracy operators in both bases are assumed to have the same
spectrum.  In one model the interaction commutes with $\hat{F}^i$
while in the other the interaction commutes with $\hat{G}^i$.  Because 
$\hat{M}$ does not commute with $\hat{M}_0$, if the relation between
the $(f,d)$ and $(g,d)$ bases involves the mass,
these two interactions cannot be the same.

Nevertheless, the form of the dynamical equation (\ref{eq:JU}) is identical 
in both cases.  Both will give the same bound state masses and 
scattering matrix elements.
It follows, using Theorem 3,  that the dynamical models constructed using the 
free particle bases
\begin{equation}
(\hat{F},\hat{d},\hat{M}_1)\qquad \mbox{and} \qquad 
(\hat{G},\hat{d},\hat{M}_2)
\label{eq:JW}
\end{equation}
are scattering equivalent and are related by 
\begin{equation}
\hat{A} = \Omega_{\pm}  (\hat{M}_1 , \hat{\Phi}_1  , \hat{M}_{\cal A})
\Omega^{\dagger} _{\pm}  (\hat{M}_2 , \hat{\Phi}_2  , \hat{M}_{\cal A}).
\label{eq:JX}
\end{equation}
The transformation $\hat{A}$ is not simply a change of basis; it is 
interaction dependent and changes the nature of the interactions.   This 
illustrates the relation of the basis choice to the structure 
of the dynamics.

To understand the nature of the interaction dependence of $\hat{A}$ note
that both wave operators in (\ref{eq:JX}) need to be computed in the 
same basis.  This leads to an expression of the form
\[
\langle f \vert \hat{A} \vert f' \rangle = \int 
\langle f \vert \Omega_{\pm}  (\hat{M}_f , \hat{\Phi}_f  , \hat{M}_{\cal A})
\vert  f'' \rangle d\mu(f'')
\langle f'' \vert g'' \rangle_{\cal A} d \mu(g'') \times
\]
\begin{equation}  
\langle g'' \vert \Omega^{\dagger} _{\pm}  
(\hat{M}_g , \hat{\Phi}_g  , \hat{M}_{\cal A}) \vert g' \rangle d\mu (g') 
\langle g' \vert f' \rangle . 
\label{eq:JXX}
\end{equation}
If the change of basis $f \leftrightarrow g$ involves the mass parametrically, 
then $\langle f \vert g \rangle_{\cal A}$ will involve the physical mass
eigenvalues while $\langle g \vert f \rangle$ involves the non-interacting 
masses.    The interaction dependence is due to having the interacting mass 
in one of these expressions and the free mass in the inverse expression.
In the limit that the interactions are turned off, this becomes the identity.

This completes the construction of the two-body dynamics.  The construction 
provides a relativistic two-body model for any choice of basis 
and $ISL(2,C)$ Clebsch-Gordan coefficient. 

To illustrate the structure of the dynamical equation
in a familiar basis 
consider the case (see Appendix II) that $\hat{F} = \lbrace 
\vec{\hat{P}} , \hat{j}_{cz}\rbrace $,
corresponding to the linear momentum and $z$-component of the
canonical spin, and $\hat{D}^i = \lbrace j_1,m_1,j_2,m_2,\hat{l},\hat{s} 
\rbrace$ where $\hat{l}$, $\hat{s}$ are
two-body orbital and spin angular momenta. The matrix elements of
$\hat{V}= \hat{M}-\hat{M}_0$ have the form:
\[
\langle  p,\mu, l,s ;m,j \vert \hat{V} \vert p',\mu',l',s' ;m',j' \rangle
=
\]
\begin{equation}
\delta (\vec{p}-\vec{p}\,') \delta_{\mu \mu'} \delta_{jj'} 
\langle l,s, m  \Vert \hat{V}^j \Vert l',s', m \rangle .
\label{eq:JP}
\end{equation}
If $m$ is replaced by the kinematic momentum $q$ defined by 
\begin{equation}
m = \sqrt{q^2 + m_1^2}+ \sqrt{q^2 + m_2^2}
\label{eq:JQ}
\end{equation}
the matrix element (\ref{eq:JP}) has the same structure as the 
corresponding non-relativistic interaction.  The eigenvalue equations
(\ref{eq:JU}) becomes: 
\[
(m - \sqrt{q^2 + m_1^2} + \sqrt{q^2 + m_2^2}) \phi_m^j ( l,s,q ) =
\]
\begin{equation}
\sum_{l'=0}^{\infty} \sum_{s'=\vert j-l\vert}^{\vert j+l\vert }  
\int_0^{\infty}  q^{\prime 2} dq' 
\langle l,s,q  \Vert \hat{V}^j \Vert l',s',q' \rangle \phi_m^j ( l',s',q' ).
\label{eq:JR}
\end{equation}

\section{The N-Body Problem}

The formulation of the $N$-body problem is by induction.  The
construction follows \cite{sokolov}\cite{fcwp}\cite{klink}.  What is
different is that the notion of ``form of the dynamics'' is replaced by a
choice $(f,d)$ of basis for $ISL(2,C)$ irreducible representation
spaces and associated Clebsch-Gordan coefficients.

The construction of the $N$-body dynamics exploits the scattering
equivalence of two representations of $ISL(2,C)$.  One representation
satisfies algebraic cluster properties and the other has a kinematic spin, 
which is useful for the $ISL(2,C)$ invariant addition of interactions.  

The construction begins with the decomposition of the system into
interacting subsystems, which are obtained by turning off the
interactions between particles in different clusters of a partition
$a$.  The tensor product of the subsystem representations define
unitary representation of $ISL(2,C)$ on the N-body Hilbert space.
These representations are reducible and have interactions in both the
N-body mass and spin operators.  As $a$ runs over all partitions these
representation contain all interactions except the $N$-body
interactions.  Because the mass and spin operators for different
decompositions into subsystems do not all commute, these tensor product
representations are not suited to $ISL(2,C)$ invariant addition of
interactions.

In order to facilitate the invariant addition of interactions,
scattering equivalences are introduced that transform each of the
tensor product representations into scattering equivalent
representations of $ISL(2,C)$ where $\hat{j}^2$, $\hat{F}^j$ and
$\Delta \hat{F}^j$ are free of interaction.  In these representations
all of the interactions are in the mass operators.  Linear combinations
of the mass operators for different decompositions into subsystems can
be used to construct an overall N-body mass operator $\bar{M}$ that
still commutes with the non-interacting operators $j_0^2$, $\hat{F}_0^j$,
and $\Delta \hat{F}_0^j$.

The existence of the required scattering equivalences follows 
by induction from 
properties of the two-body solution.  This is different than the solution 
presented in \cite{fcwp} where the kinematic subgroup and the
$\vec{p}=0$ condition 
played a 
central role in establishing the required scattering equivalences. 

The properties of $\bar{M}$ guarantee that $ISL(2,C)$ generators
expressed as functions of $\bar{M}$, $j_0^2$, $\hat{F}_0^j$, and
$\Delta \hat{F}_0^j$ satisfy the $ISL(2,C)$ commutation relations.
The associated unitary representation of $ISL(2,C)$, which is
constructed using the same method used in the two-body construction, does not
satisfy algebraic cluster properties for $N>2$.  Cluster properties
are restored by constructing a suitable scattering equivalence, which
introduces additional many-body interactions and introduces a
non-trivial interaction dependence in the spin.

The induction begins with the two-body dynamics formulated in the 
previous section.  The dynamical two-body representation,  
$\hat{U}[\Lambda ,Y]$, of $ISL(2,C)$ satisfies:

\begin{itemize} 

\item It becomes the tensor product of two one-body representations 
when the interaction is set to zero: 

\begin{equation}
\hat{U}_{(12)}[\Lambda ,Y]  \to
\hat{U}_{1} [\Lambda ,Y]  \otimes \hat{U}_{2} [\Lambda ,Y].  
\label{eq:KA}
\end{equation}

\item The two-body mass operator commutes with the 
non-interacting $\hat{F}^j$, $\Delta \hat{F}^j$ and $\hat{j}^2$:

\begin{equation}
[\hat{M}_{(12)}, \hat{F}^j_0] = [ \hat{M}_{(12)},\Delta \hat{F}^j_0
] =[\hat{M}_{(12)},\hat{j}_0^2]=0.
\label{eq:KB}
\end{equation}

\end{itemize}

These conditions cannot be simultaneously satisfied for
systems of more than two particles.   They are replaced by the 
following induction assumption, which reduces to the above 
condition when $N=2$:

\begin{itemize} 

\item For each proper subsystem $s$ of the $N$-body system, 
there is a dynamical representation $\hat{U}_s [\Lambda ,Y] :{\cal
H}_{s} \to {\cal H}_{s}$ with short-range interactions satisfying
algebraic cluster properties.  This means that if the interactions
between particles in different clusters of the subsystem $s$ are set
to zero then

\begin{equation}
\hat{U}_{s}[\Lambda ,Y]  \to \otimes_i 
\hat{U}_{a_i} [\Lambda ,Y].  
\label{eq:KC}
\end{equation}

\item For each proper subsystem there is a scattering equivalence 
$\hat{C}_{s}$ satisfying 
\begin{equation}
\hat{C}_{s} \hat{U}_{s} [\Lambda ,Y]
\hat{C}^{\dagger}_{s} = \bar{U}_{s} [\Lambda ,Y]
\label{eq:KD}
\end{equation}
with the property that the mass operator $\bar{M}_{s}$ of the 
$\bar{U}_s[\Lambda ,Y]$ representation 
commutes with $\hat{F}^i_{s}$,
$\hat{j}^2_{s}$, $\Delta \hat{F}^i_{s}$ of 
the non-interacting subsystem, $s$.

\end{itemize}  

These conditions are trivially satisfied by the two-body construction
of the previous section for $\hat{C}_s=\hat{I}$ on each 
single particle Hilbert space.

First we show that if these conditions hold for all proper subsystems then
they hold for any non-trivial partitioning of the $N$-body system.

The theorem below ensures the scattering equivalence of tensor products of 
subsystem representations that satisfy (\ref{eq:KD}) to representations 
with a non-interacting
$\hat{j}^2$, $\hat{F}^i$, and $\Delta \hat{F}^i$.

\bigskip

\noindent{\bf Theorem 4:} Let $a$ be a partition of the $N$-particle system 
into $n_a$ disjoint mutually non-interacting subsystems, $a_i$.
Assume that each subsystem has a dynamical representation
$\hat{U}_{a_i}[\Lambda ,Y]$ of $ISL(2,C)$ with an asymptotically
complete scattering theory. Assume the each of the representations
$\hat{U}_{a_i}[\Lambda ,Y]$ is scattering equivalent to a
representation that has $\hat{F}^j_{a_i} =
\hat{F}^j_{0a_i}$, $\Delta
\hat{F}^j_{a_i} =
\Delta \hat{F}^j_{0a_i}$, $\hat{j}^2_{a_i} = \hat{j}^2_{0a_i}$.
Let
\begin{equation}
\hat{U}_a[\Lambda ,Y] := \otimes_{i=1}^{n_a} \hat{U}_{a_i} [\Lambda, Y]
\label{eq:KXA}
\end{equation}
be the tensor product of subsystem representations of $ISL(2,C)$.
Then $\hat{U}_a[\Lambda ,Y]$ is scattering equivalent to a
representation $\bar{U}_a[\Lambda ,Y]$ that 
has $\hat{F}^j_{a} = \hat{F}^j_{0}$, $\Delta
\hat{F}^j_{a} = \Delta \hat{F}^j_{0}$, $\hat{j}_{a} = \hat{j}_{0}$.

This states that if the subsystem mass operators are scattering
equivalent to the subsystem mass operator with kinematic $\hat{F}^j_{a_i}$,
$\Delta \hat{F}^j_{a_i}$, $\hat{j}_{a_i}$ then the tensor product 
of the subsystems has a mass operator that is scattering equivalent 
to a mass operator with kinematic $\hat{F}^j$, $\Delta 
\hat{F}^j$, and $\hat{j}$.

The induction assumptions (\ref{eq:KD}) and (\ref{eq:KC}) and 
the application of Theorem 4 imply that for every partition $a$ 
with at least two non-empty clusters there are 
representations $\hat{U}_{a} [\Lambda ,Y]$, and
$\bar{U}_{a} [\Lambda ,Y]$, related by a scattering equivalence
$\hat{B}_a$.  The proof of Theorem 4 as well as the construction
of $\hat{B}_a$ is  given in Appendix IV.

To establish algebraic cluster properties let
$\hat{X}$ be an operator valued function of the interactions.  Assume
that a coupling constant $\lambda_b$ is put in front of all
interactions involving particles in different clusters of a partition
$b$.  Let $(\hat{X})_b$ denote the operator obtained from
$\hat{X}$ by setting $\lambda_b$ to $0$.

Theorem 4 implies the following relation:
\begin{equation}
\bar{U}_{a} [\Lambda ,Y] = \hat{B}_a \hat{U}_{a} [\Lambda ,Y]
\hat{B}^{\dagger}_a.
\label{eq:KX}
\end{equation}
Turning off interactions between
particles in different clusters of $b$ in (\ref{eq:KX})
gives, using (\ref{eq:KC}) and (\ref{eq:KXA}),  
\begin{equation}
(\bar{U}_{a} [\Lambda ,Y])_b  = (\hat{B}_a)_b \hat{U}_{a\cap b} 
[\Lambda ,Y]
(\hat{B}^{\dagger}_a)_b
\label{eq:KY}
\end{equation}
when $b \cap a$ is a refinement of $a$.  

Applying Theorem 4 directly to the partition $c=b\cap a$ gives
\begin{equation}
\bar{U}_{a\cap b } [\Lambda ,A]  = \hat{B}_{a\cap b} \hat{U}_{a\cap b} 
[\Lambda ,Y]
\hat{B}^{\dagger}_{a\cap b}.
\label{eq:KZ}
\end{equation}
This gives distinct scattering equivalences $\hat{B}_{a\cap b}$ and
$(\hat{B}_{a})_b$ relating $\hat{U}_{a\cap b} [\Lambda ,Y] $ to 
different representations that commute with 
$\hat{F}^j_{0}$, $\Delta \hat{F}^j_{0}$, and $\hat{j}^2_{0}$.
An illustration of this ambiguity in the four-body system occurs for 
$a= (123)(4)$, $b=(12)(34)$ and $c=(12)(3)(4)$.

It is desirable that the scattering equivalence obtained by turning 
off interactions agree with the scattering equivalence constructed 
directly by applying Theorem 4 to the tensor products.
This can be achieved by recursively replacing  the operators $\hat{B}_a$
of Theorem 4 with operators $\hat{A}_a$ that satisfy $(\hat{A}_a)_b = 
\hat{A}_{a\cap b}$.  This replacement involves a redefinition of 
the $\bar{M}_a$'s.

For $N-1$ cluster partitions define 
\begin{equation}
\hat{A}_a := \hat{B}_a.
\label{eq:KAB}
\end{equation}
Because $N-1$ cluster interactions only have two-body interactions,
both $\hat{A}_a$ and $\hat{B}_a$ become the identity when the 
interaction is turned off:
\begin{equation}
(\hat{A}_a)_b = \hat{A}_{a \cap b } = \hat{I}
\label{eq:KABB}
\end{equation}
In this case any non-trivial 
refinement of $a$ gives $N$ free particles.

Next consider a partition $a$ with $k$ clusters.  By induction assume
that scattering equivalences $\hat{A}_c$ have have been defined for 
all partitions $c$ with more than $k$ clusters and that these
operators satisfy $(\hat{A}_c)_d= \hat{A}_{c\cap d}$ for $n_c>k$.

Let $b$ be a partition such that $a\cap b$ has more than $k$ clusters.
Note that 
\begin{equation}
\hat{A}_{a\cap b} (\hat{B}^{\dagger}_a)_b 
\label{eq:KAA}
\end{equation}
is defined and
commutes with  $\hat{F}_{0}^{j}$, $\Delta \hat{F}_{0}^{j}$,
and $\hat{j}_{0}$.

Define
\begin{equation}
\hat{A}_a := \left ({\hat{I} - i \hat{\alpha}_a \over 
\hat{I} + i \hat{\alpha}_a} \right ) 
\hat{B}_a
\label{eq:KAC}
\end{equation}
where 
\begin{equation}
\hat{\alpha}_a := - \sum_{b\not=a} \mu (a \supset b )\hat{\alpha}_{a,b} 
\label{eq:KAD}
\end{equation}
and
\begin{equation}
\hat{\alpha}_{a,b} := i { \hat{I} - \hat{A}_b (\hat{B}^{\dagger}_a)_b 
\over \hat{I} + \hat{A}_b (\hat{B}^{\dagger}_a)_b  }.  
\label{eq:KAE}
\end{equation}
Note that $a\cap b=b$ was used in (\ref{eq:KAE}).  These expressions
utilize Cayley transforms to construct unitary functions of scattering
equivalences.  The resulting unitary operators will be scattering
equivalences provided their Cayley transforms are in the algebra of 
asymptotic constants.  This is not entirely trivial, because the 
algebra ${\cal C}$ is uniformly closed, but not strongly closed.  $\hat{A}_a$ 
will be a scattering equivalence if the Cayley transforms 
$\hat{\alpha}_{a,b}$ are bounded. This will be assumed in all that follows.

The restriction $b\not=a$ means that the $b$'s appearing in the sum are 
proper refinements of $a$ and necessarily have more than $k$ clusters. 
By induction the
$\hat{A}_b$'s satisfy $(\hat{A}_b)_c = \hat{A}_{b \cap c}$. It follows 
for $c\cap a \not= a$ that
\begin{equation}
(\hat{\alpha}_{a,b})_c := i { \hat{I} -  \hat{A}_{b\cap c} 
(\hat{B}^{\dagger}_a)_{b \cap c}  
\over \hat{I} + \hat{A}_{b\cap c} (\hat{B}^{\dagger}_a)_{b\cap c} }  =
\hat{\alpha}_{a, b \cap c} 
\label{eq:KAF}
\end{equation}
which gives
\[
(\hat{\alpha}_a)_c = - \sum_{b\not=a} \mu (a \supset b )
\hat{\alpha}_{a,b\cap c} = 
\]
\begin{equation}
- \sum_{b\not=a} \mu (a \supset b )\zeta ((b \cap c)\supset d) 
\mu (d \supset e) \hat{\alpha}_{a,e} . 
\label{eq:KAG}
\end{equation}
Using (\ref{eq:ID}) gives
\begin{equation}
- \sum_{b\not=a} \mu (a \supset b )\zeta (b \supset d)
\zeta (c \supset d) \mu (d \supset e) \hat{\alpha}_{a,e}.
\label{eq:KAH}
\end{equation}
The $b$ sum gives $\mu (a \supset a)\zeta(a \supset d) -\delta_{ad}= 
\zeta(a \supset d)-\delta_{ad}$.
Using this in the above sum and observing that $\zeta (c \supset a) =0$,
gives
\begin{equation}
\sum_{d,e} \zeta (a \supset d)
\zeta (c \supset d) \mu (d \supset e) \hat{\alpha}_{a,e}
=
\sum_{d,e} \zeta (a\cap c \supset d) \mu (d \supset e) \hat{\alpha}_{a,e}
= \hat{\alpha}_{a, a\cap c}.
\label{eq:KAHA}
\end{equation}

It follows that
\[
(\hat{A}_a)_c = 
({\hat{I} - i \hat{\alpha}_{a, c} \over \hat{I} + i \hat{\alpha}_{a, c}}) 
(\hat{B}_a)_c =
\]
\begin{equation}
\hat{A}_{a \cap c} (\hat{B}^{\dagger}_a)_c (\hat{B}_a)_c = \hat{A}_{a \cap c} .
\label{eq:KAI}
\end{equation}
This shows that if the result holds for more than $k$ clusters, it holds for
$k$ clusters.

This process can be continued recursively until $n_a=2$.
The result is a set of scattering equivalences, $\hat{A}_a$  and 
representations
\begin{equation}
\hat{U}_a[\Lambda, Y], \bar{U}_a[\Lambda, Y]
\label{eq:KAJ}
\end{equation}
with the properties 
\begin{equation}
\bar{U}_a[\Lambda, Y]= \hat{A}_a \hat{U}_a[\Lambda, Y]\hat{A}^{\dagger}_a
\label{eq:KAK}
\end{equation}
\begin{equation}
\hat{U}_a[\Lambda, Y]= \otimes_{i=1}^{n_a} \hat{U}_{a_i}[\Lambda, Y]
\label{eq:KAL}
\end{equation}
\begin{equation}
\hat{A}_a \to \hat{A}_{a\cap b} 
\label{eq:KAM}
\end{equation}
and 
\begin{equation}
\bar{F}_{a}^i =
\hat{F}_{0}^i,\qquad 
\Delta \bar{F}_{a}^i =
\Delta \hat{F}_{0}^i, \qquad 
\bar{j}^2_{a} = \hat{j}^2_{0}.
\label{eq:KAN}
\end{equation} 

The final step is to complete the construction of the dynamics.
For each partition $a$ of the $N$-particle system with at least two 
clusters let $\hat{M}_a$ be the mass operator for the tensor product 
representation $\hat{U}_a[\Lambda ,Y]$.  Note that 
\begin{equation}
\bar{M}_a = \hat{A}_a \hat{M}_a \hat{A}_a^{\dagger} 
\label{eq:KAO}
\end{equation}
is scattering equivalent to $\hat{M}_a$ and commutes with 
$\hat{F}_0^{j}$, $\Delta \hat{F}_{0}^j$, and $\hat{j}^2_0$.

Define
\begin{equation}
\bar{M} := - \sum_{a \not= 1}  \mu (1 \supseteq a ) \bar{M}_a + [\bar{M}]_N =  
- \sum_{a \not= 1}  \mu (1 \supseteq a )  
\hat{A}_a \hat{M}_a \hat{A}_a^{\dagger} + [\bar{M}]_N  
\label{eq:KAP}
\end{equation}
where $[\bar{M}]_N$ is a possible additional $N$-body interaction that 
commutes $\hat{F}^j_{0}$, $\Delta \hat{F}^j_{0}$, and $\hat{j}^2_0$.
By construction $\bar{M}$ commutes with 
$\hat{F}^j_{0}$, $\Delta \hat{F}^j_{0}$, and $\hat{j}^2_0$.  This 
expansion is equivalent to the cluster expansion of $\bar{M}$. 
By the induction assumption, turning off the interactions between particles in 
different clusters of partition $b$ gives
\[
(\bar{M})_b  := - \sum_{a \not= 1}  \mu (1 \supseteq a ) (\bar{M}_a)_b =  
- \sum_{a \not= 1}  \mu (1 \supseteq a )  
\hat{A}_{a\cap b} \hat{M}_{a\cap b}  \hat{A}_{a\cap b}^{\dagger}=  
\]
\[
- \sum_{a \not= 1} \mu (1 \supseteq a)  \zeta ((a \cap b) \supseteq d) 
\mu (d \supseteq e ) 
\hat{A}_{e} \hat{M}_{e}  \hat{A}_{e}^{\dagger}=  
\]
\begin{equation}
- \sum_{a \not= 1}  \mu (1 \supseteq a ) \zeta (a \supseteq d)
\zeta (b \supseteq d) \mu (d \supseteq e ) 
\hat{A}_{e} \hat{M}_{e}  \hat{A}_{e}^{\dagger}.
\label{eq:KAPA}
\end{equation}
The $a$ sum gives $(1-\delta_{1 d})\hat{I} $.  Inserting this into 
(\ref{eq:KAPA}) gives
\begin{equation} 
(\bar{M})_{b} =\hat{A}_{b} \hat{M}_{b}  \hat{A}_{b}^{\dagger}
\label{eq:KAQ}
\end{equation}
or
\begin{equation}
(\bar{M})_b = \bar{M}_b.
\label{eq:KAR}
\end{equation}

This is not the mass operator $\hat{M}_{b}$ corresponding to the tensor 
product 
of the subsystems associated with the clusters of $b$.  To correct this 
define the scattering equivalence
\begin{equation}
\hat{A}:= {I + i \hat{\alpha} \over I -i \hat{\alpha} }
\label{eq:KAS}
\end{equation}
with
\begin{equation}
\hat{\alpha} = - \sum_{a\not= 1} \mu (1 \supseteq a) \hat{\alpha}_a
\label{eq:KAT}
\end{equation}
\begin{equation}
\hat{\alpha}_a := i{\hat{I} - \hat{A}_a \over \hat{I}+ \hat{A}_a}.
\label{eq:KAU}
\end{equation}
Using the same algebra used to show that $(\bar{M})_b=\bar{M}_b$
it follows that 
\begin{equation}
(\hat{A})_b = \hat{A}_b.
\label{eq:KAV}
\end{equation}
Since $\hat{A}$ is a scattering equivalence define
\begin{equation}
\hat{M}:= \hat{A}^{\dagger} \bar{M} \hat{A}.  
\label{eq:KAW}
\end{equation}

Since $\bar{M}$ commutes with the kinematic operators
$\hat{F}_{0}^j$, $\Delta \hat{F}_{0}^j$, and $\hat{j}^2_0$, simultaneous 
eigenstates of $\bar{M}$, $\hat{F}_{0}^j$, and $\hat{j}^2_0$ define a complete
set of states states that transform irreducibly.  This can be used
to construct a representation $\bar{U}[\Lambda ,Y]$ of the $ISL(2,C)$.
The scattering equivalence $\hat{A}$
defines a scattering equivalent representation 
\begin{equation}
\hat{U}[\Lambda ,Y]:= \hat{A}^{\dagger} \bar{U}[\Lambda, Y] \hat{A} 
\label{eq:KAX}
\end{equation}
with the property that 
\begin{equation}
(\hat{U}[\Lambda ,Y])_b := \hat{A}_b^{\dagger} \bar{U}_b[\Lambda, Y] \hat{A}_b 
= \hat{U}_b [\Lambda ,Y]= \otimes_{i=1}^{n_b} \hat{U}_{b_i} [\Lambda ,Y].
\label{eq:KAY}
\end{equation}
The generators have the form
\begin{equation}
\hat{P}^{\mu} =  \hat{A}^{\dagger}P^{\mu}(\bar{M},\hat{j}_0^2, 
\hat{F}_0^i, \Delta \hat{F}_0^i )   \hat{A}
\end{equation}
and 
\begin{equation}
\hat{J}^{\mu\nu } =  \hat{A}^{\dagger}J^{\mu\nu}(\bar{M},\hat{j}_0^2, 
\hat{F}_0^i, \Delta \hat{F}_0^i )   \hat{A} .
\end{equation}
This completes the proof of the induction. 

The operator $\hat{U}[\Lambda ,Y]$ defined in (\ref{eq:KAX}) is the
desired $N$-body representation of $ISL(2,C)$ that is consistent with
the dynamics and satisfies algebraic cluster separability.  The effect
of the transformation $\hat{A}$ is to cancel the $\hat{A}_a$'s from
the subsystems.  It generates new many-body interactions
that are necessary for the algebraic cluster properties of 
$\hat{U}[\Lambda ,Y]$.

To summarize this construction; tensor products
of subsystem dynamics are transformed to scattering equivalent
representations where the operators $\hat{F}^j, \Delta
\hat{F}^j$, and $\hat{j}$ are free of interactions.  The 
transformed mass operators are combined to construct a mass operator
for a unitary representation of $ISL(2,C)$ with kinematic $\hat{F}^j,
\Delta
\hat{F}^j$, and $\hat{j}$.  This representation is transformed to a 
scattering equivalent representation satisfying cluster properties.

The construction, while complex, leads to a simple structure.  All of
the $ISL(2,C)$ generators can be expressed as sums of one, two, three,
$\cdots$, N-body interactions.  For any $ISL(2,C)$ generator, the
$k$-body interaction in the $k$-body problem is identical to the
$k$-body interaction in the many-body problem.  At each stage of the
construction the subsystem interactions remain unchanged. What is new
is that cluster properties generate new many-body interactions.  These
do not change when they are imbedded in systems with more than $N$
particles.  The spin, which is a non-linear function of these
generators, is an interaction dependent quantity given by 
\begin{equation}
\hat{j}^2 =  \hat{A}^{\dagger}\hat{j}_0^2 \hat{A} .
\end{equation} 
The scattering equivalence $\hat{A}$ is an interaction dependent operator 
that becomes the identity when the interactions are switched off.
While there is
freedom to include many-body interactions, there is a class of
many-body interactions that cannot be removed without violating
cluster properties.

\section{Cluster Equivalence}

The dynamical unitary representation of $ISL(2,C)$ constructed in the
previous section satisfies algebraic cluster properties.  With
suitable short ranged interactions it will satisfy cluster properties
(\ref{eq:HB}) and the spectral condition.  The choice of basis $(f,d)$
was an important element of this construction.  In this section, this
representation is shown to be scattering equivalent to a
representation based on a different choice of basis, $(g,h)$.  This
representation also satisfies algebraic cluster properties.

This illustrates the existence of a subgroup of the group of
scattering equivalences that relates the constructions based on
different irreducible representation basis choices and preserves
algebraic cluster properties.  This subgroup will be called the 
group of cluster equivalences.

It follows that the choice of irreducible basis used in the
construction has no fundamental physical significance.  This
generalizes the equivalence of choices of kinematic subgroups in
two ways.  First, it extends the result to the general setting of this
paper where the form of dynamics is replaced by the basis choice
$(f,d)$.  Second, it shows that this equivalence respects cluster
properties.

To illustrate the nature of the required scattering equivalence first let
$\hat{U}^{f}[\Lambda ,Y]$ denote the representation constructed in the 
previous section using the $(f,d)$ basis.  Turning off interactions 
between particles in different clusters of the partition $a$ gives
\begin{equation}
\hat{U}^{f}[\Lambda ,Y] \to \hat{U}_a^{f}[\Lambda ,Y] =
\hat{A}_a^{f\dagger} \bar{U}_a^{f}[\Lambda ,Y] \hat{A}_a^f 
\end{equation}
where $\hat{A}^f_a$ are the scattering equivalences constructed in the 
previous section.  The superscript $f$ indicates that the $(f,d)$
basis was used in the construction. 

Algebraic cluster properties give the relations
\[
\hat{U}^{f}[\Lambda ,Y] \to \hat{U}_a^{f}[\Lambda ,Y] =
\otimes_{i=1}^{n_a} \hat{U}_{a_i}^{f}[\Lambda ,Y] =
\]
\begin{equation}
\otimes_{i=1}^{n_a}\left 
( \hat{A}_{a_i}^{f\dagger} \bar{U}_{a_i}^{f}[\Lambda ,Y]
\hat{A}_{a_i}^{f} \right ) =
( \otimes_{i=1}^{n_a}
\hat{A}_{a_i}^{f\dagger}) (\otimes_{i=1}^{n_a} \bar{U}_{a_i}^{f}[\Lambda ,Y]
)(
\otimes_{i=1}^{n_a}\hat{A}_{a_i}^{f}) 
\end{equation}
where the $\hat{A}^f_{a_i}$ are the $\hat{A}^f$ operators for the subsystem 
consisting of the particles in the $i-th$ cluster of $a$.

It is useful to introduce the operators 
\begin{equation} 
\tilde{U}_a^f [\Lambda ,Y]
:= \otimes_{i=1}^{n_a} \bar{U}_{a_i}^{f}[\Lambda ,Y]
\end{equation}
which are related to $\hat{U}^f_a[\Lambda ,Y]$ by the scattering equivalence 
\begin{equation}
\hat{B}^f_a := \otimes_{i=1}^{n_a}\hat{A}_{a_i}^{f}.
\end{equation}
The construction of the previous section defined $\hat{U}^f_a
[\Lambda,Y]:= \tilde{U}^f_a [\Lambda ,Y]$ for $n_a=N-1$.  All of the
$\hat{U}_a^f [\Lambda ,Y]$'s were recursively constructed from the
$n_a=N-1$ cluster representations.

Any of the representations $\bar{U}^f_a [\Lambda ,Y]$
are scattering equivalent to a $\bar{U}^g_a [\Lambda ,Y]$
representation.  This scattering equivalence is realized by 
making the following replacements in the kernel of the 
barred mass operators:
\begin{equation}
\langle f,d(m_0,j_0 ) \vert \bar{M}^f \vert f',d'(m_0',j_0' )
\rangle = 
\delta [f;f'] \delta_{j_0,j_0'} \langle m_0,d \Vert \bar{M}^{j_0} 
\Vert m_0',d' \rangle
\end{equation}
by
\begin{equation}
\langle g,h(m_0,j_0 ) \vert \bar{M}^g \vert g',h'(m_0',j_0' )
\rangle = 
\delta [g;g'] \delta_{j_0,j_0'} \langle m_0,h \Vert \bar{M}^{j_0} 
\Vert m_0',h' \rangle
\end{equation}
where the reduced kernel $\langle m_0,h \Vert \bar{M}^{j_0} \Vert
m_0',h' \rangle$ is defined in terms of the reduced kernel 
$ \langle m_0,d \Vert \bar{M}^{j_0} \Vert m_0',d'
\rangle$ by a variable change $d \to h$ implemented by
kinematic $ISL(2,C)$-Racah coefficients.  This means abstract 
reduced mass operators are identical.
The operators $\bar{M}^g$ and $\bar{M}^f$ differ because of the
delta functions in $f$ or $g$; but both operators   
manifestly give the same $S$ matrix elements and bound-state observables.
The operators $\bar{M}^f$ and $\bar{M}^g$ define
scattering equivalent representations of $ISL(2,C)$ with the 
non-interacting $\hat{F}^i$, $\Delta \hat{F}^i$ or $\hat{G}^i$, 
$\Delta \hat{G}^i$ respectively.  The scattering equivalence is 
denoted by $\hat{C}^{gf}$:
\begin{equation}
\hat{C}^{gf} \bar{U}^f [\Lambda ,Y] \bar{C}^{gf\dagger} = 
\bar{U}^g [\Lambda ,Y]
\label{eq:LAA}
\end{equation}

Since this equivalence is valid for systems or subsystems, for each 
partition $a$ the following representations are scattering equivalent:
\begin{equation}
\hat{U}_a^f [\Lambda ,Y],\bar{U}_a^f [\Lambda ,Y],\tilde{U}_a^f [\Lambda ,Y],
\bar{U}_a^g [\Lambda ,Y],\tilde{U}_a^g [\Lambda ,Y].
\end{equation}

These representations have the property that 
$\hat{U}_a^f [\Lambda ,Y]=\tilde{U}_a^f [\Lambda ,Y]$ for $N-1$ cluster
partitions and 
$\hat{U}_a^f [\Lambda ,Y]$ is scattering equivalent to 
$\bar{U}_a^f [\Lambda ,Y]$ for the $1$-cluster 
partition. 

The goal is to find  a $\hat{U}^g[\Lambda ,a]$ that is scattering equivalent 
to $\bar{U}^g [\Lambda ,Y]$ and $\bar{U}^f[\Lambda, Y]$ and also satisfies 
algebraic cluster properties, with 
$\hat{U}^g_a [\Lambda ,Y] = \tilde{U}^g_a [\Lambda ,Y]$ for $n_a= N-1$. 

The first step is to define 
\begin{equation}
\hat{U}^g_a [\Lambda ,Y] = \tilde{U}^g_a [\Lambda ,Y]
\end{equation}
for $n_a= N-1$.  Following the construction of the previous 
section, this gives scattering equivalences $\hat{A}_a^g$ 
relating $\hat{U}^g_a [\Lambda ,Y]$ to $\bar{U}^g_a [\Lambda ,Y]$
for $n_a=N-1$. 

Next, assume by induction that $\hat{U}^g_a [\Lambda ,Y]$ has been
defined for partitions with more than $K$ clusters satisfying
algebraic cluster properties and is scattering equivalent to
$\bar{U}^g_a [\Lambda ,Y]$.  The $\bar{U}^g_a [\Lambda ,Y]$ for
$K$-cluster partitions is defined by (\ref{eq:LAA}).  Its mass operator,
$\bar{M}^g_a$, is related to $\bar{M}^f_a$ by replacing
delta functions in $f$ by delta functions in $g$.  Since
$(\bar{M}^f_a)_b= \bar{M}^f_{a\cap b}$ it follows that $(\bar{M}^g_a)_b=
\bar{M}^g_{a\cap b}$ because the kernel of the two operators only differ 
by delta functions in the overall kinematic operators $f$ or $g$. 

This means that $\bar{M}^g_a$ differs from the cluster expansion 
\begin{equation}
\bar{M}^{g0}_a = - \sum_{b \not= a} \mu (a \supseteq b) \bar{M}^g_b
\end{equation}
by at most an $a$-connected interaction term, $[\bar{M}]_a^g$.  
In order to construct 
the desired representation it is enough to define 
\begin{equation}
\hat{U}^g_a[\Lambda ,Y] := \hat{A}^{g\dagger}_a \bar{U}^g_a [\Lambda ,Y] 
\hat{A}^{g}_{a}  
\end{equation}
where
\begin{equation}
\hat{A}^g_a = {I + i \hat{\alpha}^g_a \over I - i \hat{\alpha}^g_a}
\end{equation}
\begin{equation}
\hat{\alpha}^g_a := - \sum_{b \not= a} \mu (a,b) \hat{\alpha}^g_b
\end{equation}
\begin{equation}
\hat{\alpha}_b^g = i { I + \hat{A}^g_b \over
I - \hat{A}^g_b }. 
\end{equation}
Following the algebra used in (\ref{eq:KAPA}) $\hat{\alpha}^g_a$ 
has the property that 
\begin{equation}
(\hat{\alpha}^g_a)_b = \hat{\alpha}^g_b \qquad b \subset a
\end{equation}
and
\begin{equation}
(\hat{U}^g_a)_b[\Lambda ,Y]  := \hat{A}^{g\dagger}_{a\cap b}  
\bar{U}^g_{a\cap p} 
[\Lambda ,Y] \hat{A}^{g}_{a \cap b} =   
\hat{U}^g_{a\cap b}[\Lambda ,Y]
\end{equation}
This differs from the result of a direct construction in the $(g,h)$
basis because of the difference $[\bar{M}]_a^g$ between $\bar{M}^g_a$ and
$\bar{M}^{g0}_a$.  This introduces additional many-body interactions
that are needed maintain the scattering equivalence at each stage of
the recursion.  Note that in this construction the factor $\mu (a
\supseteq b)$ ensures that only the $b$ satisfying $b \subset a$
appear in the sum.  These partitions have more than $K$-clusters.
This construction can be continued until $K=1$, where 
\begin{equation}
\hat{U}^g
[\Lambda ,Y]=\hat{U}^g_1 [\Lambda ,Y]= \hat{A}^g \bar{U}^g_1
\hat{A}^{g\dagger} 
\end{equation} 
is the desired representation
based on the $(g,h)$ representation.  The relevant scattering
equivalence is
\begin{equation}
\hat{U}^g[\Lambda ,Y] = \hat{A}^{g\dagger}\hat{C}^{gf}\hat{A}^f   
\hat{U}^f[\Lambda ,Y]\hat{A}^{f\dagger}\hat{C}^{gf\dagger}\hat{A}^g .
\end{equation}
It follows that $\hat{A}^{g\dagger}\hat{C}^{gf}\hat{A}^f$ is the
desired scattering equivalence connecting the construction of $\hat{U}
[\Lambda ,Y]$ using the $(f,d)$ representation to a dynamics
satisfying cluster properties based on the $(g,h)$ representation.

It is important to emphasize that the $\hat{A}^g$ constructed in this manner 
are not identical to the corresponding operators that would have been 
constructed if one began with the $(g,h)$ basis.   This is due to the 
presence of additional many-body interactions that are determined by the 
difference between the operators $\bar{M}^{g0}_a$ and $\bar{M}^{g}_a$
for each $a$.  These differences account for the dynamical differences 
that occur when the many-body dynamics is formulated with different 
basis choices, or using different forms of dynamics.

The cluster equivalences transform $ISL(2,C)$ generators in one
representation to physically equivalent generators in another
representation.  In each representation the interactions are
distributed differently among the generators.  Specific representation
have computational advantages.
 
\section{Summary and Conclusion}

This paper provides a general construction of a unitary representation
$\hat{U}[\Lambda ,Y]$ of $ISL(2,C)$ for a system of N-interacting
particles based on the representation theory of $ISL(2,C)$.  For
suitable interactions the representation satisfies cluster properties
and the spectral condition.  The representation defines a non-trivial
relativistic quantum theory of interacting particles.  Unitary
operators that preserve the S-matrix and cluster properties,
called cluster equivalences,  relate the different constructions.

Relativistic quantum theory of $N$-particles can be applied to model
systems of strongly interacting particles.  This framework has many 
features of non-relativistic quantum mechanics and local
relativistic quantum field theory.  Like non-relativistic quantum
mechanics, it is a mathematically well behaved theory where exact
numerical calculations are possible.  Like quantum field theory, it is
a quantum theory with an exact $ISL(2,C)$ symmetry that satisfies
cluster properties and the spectral condition.

The generality of the construction suggests that any quantum theory
dominated by a finite number of particle degrees of freedom which is
consistent with Poincar\'e invariance, cluster properties, and the
spectral condition will be related to a theory of the type discussed
in this paper by a cluster equivalence.

The cluster equivalences introduced in section 13 relate physically
equivalent representations of the same model.  Cluster equivalent
models have the same bound state observables and S-matrix elements.
In each representation free particles are represented as tensor
products of irreducible representations.  The unitary representation
of $ISL(2,C)$ that defines the dynamics clusters into tensor products
of subsystems representations with the same properties.  Cluster
equivalence is a stronger condition than unitary equivalence or
scattering equivalence.  Scattering equivalences were shown to be
unitary elements of the $C^*$ algebra of asymptotic constants.
Cluster equivalences were shown to be a subgroup of the scattering
equivalences.

The practical need to understand the relationship between different
formulations of relativistic quantum theory suggests that it would be
useful to have an abstract definition of a relativistic quantum theory
of particles.  The situation is different than the quantum field
theory case, were there are several sets of axioms \cite{haag} that
are designed to define a suitable local field theory, with an absence
of examples of realistic theories consistent with these axioms.  In
relativistic quantum theory there are many applications that claim to
be relativistic quantum theories, with no universally accepted
criteria of what it means to be a relativistic quantum theory of
particles.  The absence of an acceptable definition of what
constitutes a relativistic quantum theory of particles makes
comparison difficult.  The construction of this paper, which focuses
on mathematical formulation of observable physical properties, and how
they can be realized in models, suggest minimal elements that need to
be included in a set of axioms:

\begin{itemize} 

\item[A1]: The Hilbert space ${\cal H}$ is the tensor product of 
irreducible representation spaces of $ISL(2,C)$ associated 
with the mass and spins of the constituent particles.

\item[A2]: There is a unitary representation $\hat{U}[\Lambda ,Y]$ of 
$ISL(2,C)$ on ${\cal H}$ with a positive mass and energy 
spectrum.

\item[A3]: The Hilbert space can be factored into a tensor product 
of subsystem spaces,  with each one supporting a subsystem unitary 
representation $\hat{U}_i[\Lambda ,Y]$ of $ISL(2,C)$.  For
each partition $a$ into subsystems $a_i$ the operator 
$\hat{U}[\Lambda ,Y]$  satisfies cluster property 
(\ref{eq:HB})

\item[A4] The dynamics $\hat{U}[\Lambda ,Y]$ has an asymptotically complete,
$ISL(2,C)$ invariant $S$-matrix.

\end{itemize} 

These requirements can be used to formulate a precise relationship
between different formulations of relativistic quantum theory when they are
applied to systems with finite energy and number of degrees of freedom.

The construction in section 12 points to some of the general features
of relativistic quantum theory of particles.  In the physical
representations of $ISL(2,C)$ the scattering equivalence $\hat{A}$,
which is an interaction dependent operator, normally generates
interaction dependent terms in all of the operators using the 
relations:

\begin{equation}
\hat{F}^i = \hat{A}^{\dagger} \hat{F}_0^i \hat{A}
\end{equation}

\begin{equation}
\Delta \hat{F}^i = \hat{A}^{\dagger} \Delta \hat{F}_0^i \hat{A}
\end{equation}

\begin{equation}
j^2 = \hat{A}^{\dagger} j^2_0 \hat{A}
\end{equation}

\begin{equation}
P^{\mu} = P^{\mu} (M, j^2, \hat{F}^i ,\Delta \hat{F}^i) =
\hat{A}^{\dagger} 
P^{\mu} (\bar{M}, j_0^2, \hat{F}_0^i ,\Delta \hat{F}_0^i) \hat{A}
\end{equation}

\begin{equation}
J^{\mu \nu} = J^{\mu \nu} (M, j^2, \hat{F}^i ,\Delta \hat{F}^i) =
\hat{A}^{\dagger} 
J^{\mu \nu} (\bar{M}, j_0^2, \hat{F}_0^i ,\Delta \hat{F}_0^i) \hat{A}
\end{equation}
While the construction begins with representations having 
kinematic $j^2$, $\hat{F}^i$ , and $\Delta \hat{F}^i$, all of these
operators acquire an interaction dependence in the physical representation.

Tensor and spinor operator densities also play an important role in 
relativistic quantum mechanics.  For example, the hadronic electroweak 
current operators provide the coupling of the hadronic dynamics to 
electroweak probes.   In one-boson exchange approximations these
current operators must transform as 4-vector densities with respect to 
$ISL(2,C)$
\begin{equation}
\hat{U}[\Lambda ,Y] I^{\mu} [X] \hat{U}^{\dagger} [\Lambda ,Y] =
I^{\nu} [\Lambda X \Lambda^{\dagger}+Y ] \Lambda_{\nu}{}^{\mu}. 
\label{eq:RA}
\end{equation}
Because $\hat{U}[\Lambda ,Y]$ is an interaction dependent operator,
the covariance condition (\ref{eq:RA}) requires the existence 
of many-body contributions to the current.  

This is understood by considering covariance condition 
\[
\langle f;m,j \vert I^{\mu} [X] \vert f' ;m',j' \rangle = 
\]
\[
\int d\mu(f'')d\mu(f''') 
\langle f'';m,j \vert I^{\nu} [\Lambda X \Lambda^{\dagger}+Y ]
\vert f''' ;m',j' \rangle  \times 
\]
\begin{equation} 
{\cal D}^{*m,j}_{f'',f}[\Lambda,Y]
{\cal D}^{m',j'}_{f''',f'}[\Lambda,Y] \Lambda_{\nu}{}^{\mu}. 
\end{equation}
In this expression the $m$ and $m'$ in the ${\cal D}$ functions are
physical mass eigenvalues.  This expression fixes a general matrix
elements in terms of a set of independent current matrix elements and
interaction $(m)$ dependent coefficients.  This is essentially the
Wigner-Eckart theorem for $ISL(2,C)$.  In this interpretation the
interaction dependence arises because the Clebsch-Gordan coefficients
depend on the physical mass eigenvalues.  This means that the operators
$\hat{I}^{\mu}(X)$ necessarily have interaction dependent
terms that depend on the specific representation.  

The result is that the representation of tensor and spinor densities
is related to the representation of the dynamics.  Changing the
representation of the dynamics by a cluster equivalence changes the
representation of the interaction dependent parts of the tensor and
spinor densities.  This has important implications for modeling
electromagnetic probes of hadronic systems.

Dirac's forms of dynamics are obtained for special basis choices.
Specifically, if the $ISL(2,C)$ Wigner ${\cal D}$ functions, ${\cal
D}^{m,j}_{f,f'} [\Lambda,Y]$, do not depend explicitly on $m$ for a
subgroup ${\cal G}$ of $ISL(2,C)$, there are no interactions in the
generators of the subgroup.  This depends on the choice of
commuting operators $\hat{F}^i$ that are used to label vectors 
in $ISL(2,C)$ irreducible subspaces.  Cluster equivalences can be used
to relate a general model to an equivalent models in any of Dirac's 
forms of dynamics.

The author would like to thank F. Coester for critically reading this
manuscript and would like to acknowledge the generous support of the 
US Department of Energy which supported this work under 
contract \# DE-FG02-86ER40286.    

\section{Appendix I}

Examples of positive mass positive energy irreducible representations
of $ISL(2,C)$ are constructed.  The construction presented below is not
as general as the abstract construction given in section 5, but it is
general enough to include all of the representations that are commonly
used in the literature.

Let $f^i (\vec{p},m )$, $i=1,2,3$  be three 
independent real valued functions of the
three momentum and the mass.  Since the $\hat{M}$ and $\hat{\vec{P}}$ 
commute, these three functions become 
commuting self-adjoint operators when $m$ and $\vec{p}$ are replaced by 
operators. Independence means that these functions can be uniquely inverted
to express $\vec{p} = \vec{P}(f,m)$ where $f$ denotes the three functions 
$f^i$.  By the implicit function theorem this will be true provided 
the Jacobian matrix
\begin{equation}
{\partial f^i \over \partial p^j} 
\end{equation}
is invertible for any $\vec{p}$ and any $m$ in the spectrum of $\hat{M}$.  

Define the operators
\begin{equation}
\hat{F}^i = f^i(\hat{\vec{P}},\hat{M})
\end{equation}
for $i=1$ to 3.
Let $L(p)$ be an arbitrary but fixed 
$SL(2,C)$ valued function of $p=(\sqrt{m^2+ \vec{p}\,^2},\vec{p}\,)$ with 
properties 
\begin{equation}
L(p) L^{\dagger} (p) = {1 \over m} \sigma_{\mu} p^{\mu}   
\end{equation}
\begin{equation}
L(p_0) L^{\dagger} (p_0) = \sigma_{0} \qquad p_0 := (m,0,0,0).
\end{equation}
These equations mean that $L(p)$ is an $SL(2,C)$ representation of a 
Lorentz boost.  In general it can differ from the 
canonical (rotationless boost) 
by a $p$-dependent rotation, $R(p) \in SU(2)$:
\begin{equation}
L(p) = L_c (p) R(p) \qquad R(p_0) =I.
\end{equation}
Given the function $L(p)$ it is possible to define the $SL(2,C)$ valued 
matrix of operators $L(\hat{P})$ which is obtained by 
replacing $p$  by the commuting operators $(\hat{\vec{P}},
\hat{M})$.

For a given $L(p)$ define the $l$-spin by
\begin{equation}
\hat{\vec{j}}_l := {1 \over 2\hat{M}}\mbox{Tr}\left [\vec{\sigma} L(\hat{P}) 
\hat{W}^{\mu} \sigma_{\mu}
L^{\dagger}(\hat{P}) \right ] 
\label{eq:APAA}
\end{equation}
where $\hat{W}^{\mu}$ is the Pauli Lubanski vector.  Since
$\hat{W}^{\mu}$ commutes with $\hat{P}^{\nu}$, all components of
$\hat{\vec{j}}_l$ commute with $\hat{F}^1,\hat{F}^2,\hat{F}^3$.  In
addition, for any choice of $L(p)$ the components of $\vec{\hat{j}}_l$ 
satisfy $SU(2)$
commutation relations with $\hat{j}_l^2 = \hat{j}^2= \hat{W}^2 / \hat{M}^2$.  
Let $\hat{F}^4 =
\hat{z} \cdot
\hat{\vec{j}}_l$.
The operators $\hat{F}^1, \cdots \hat{F}^4,
\hat{M}, \hat{j}^2$ define a complete set of commuting self-adjoint
operators.

Let $f_0^1 = f^1 (p_0), f_0^2 = f^2 (p_0), f_0^3 = f^3 (p_0)$.  By
construction $f_0^1,f_0^2,f_0^3$ is invariant under rotations,
although $f$ does not transform like an $SO(3)$ vector.  Let
$\vec{f}$ denote the eigenvalues of $\hat{F}^1,\hat{F}^2$ and
$\hat{F}^3$ and $\mu$ denote the eigenvalue of $\hat{F}^4$.  Define
$f_\Lambda := f( \vec{p}_\Lambda ,m)$, where $p^{\mu}_\Lambda =
\Lambda^{\mu}{}_{\nu} p^{\nu}$.  For fixed $\Lambda$, $f_\Lambda$ is a 
function of $f$ and $m$.  

Let $\vert f_0,\mu ;j,m \rangle$ denote a rest eigenstate
of $\hat{F}^i, \hat{M}, \hat{j}^2$
and let $R$ be a $SU(2)$ rotation.   Define rotations and translations on
the rest states by:
\begin{equation}
\hat{U}[R,0] \vert f_0,\mu ;j,m \rangle :=
\sum_{\nu=-j}^j \vert f_0,\nu ;j,m \rangle D^j_{\nu \mu} (R) 
\label{eq:APAB}
\end{equation} 
\begin{equation}
\hat{U}[I,Y] \vert f_0,\mu ;j,m \rangle := e^{-im y^0} 
\vert f_0,\mu ;j,m \rangle .
\label{eq:APAD}
\end{equation}
Define states of arbitrary $\hat{F}$ by
\begin{equation}
\vert f,\mu ;j,m \rangle := \hat{U}[L(f),0] \vert f_0,\mu ;j,m \rangle
\sqrt{\vert {\partial f_0 \over \partial f} \vert }.
\label{eq:APAC}
\end{equation}

The expressions for $\hat{U}[R,0]$ and $\hat{U}[I,Y]$ are manifestly
unitary.  The factor $\sqrt{\vert {\partial f_0 \over \partial f}
\vert }$ assures that $\hat{U}[L(f),0]$ unitarity for states with a
delta-function normalization.  These elementary relations determine a
unitary representation $\hat{U}[\Lambda ,Y]$ on any state by using the
decomposition
\begin{equation}
\hat{U}[\Lambda ,Y]=\hat{U}[I,Y] \hat{U}[L(f_{\Lambda}),0]
\hat{U}[R_{wl} (\Lambda ,f) ,0]\hat{U}[L^{-1}(f),0]
\end{equation}
where
\begin{equation}
R_{wl} (\Lambda ,f) := L^{-1} (f_{\Lambda})\Lambda L (f)
\end{equation}
is the $l$-spin Wigner rotation and $L(f)$ is obtained from $L(p)$ 
by replacing $p$ by $p(f,m)$.

The irreducible representation in this basis follows as a consequence 
of the above relations:
\begin{equation}
\hat{U}[\Lambda ,Y] \vert f, \mu ;j,m \rangle =
\sum_{\nu=-j}^j 
\vert f_\Lambda, \nu ;j,m \rangle
e^{i p(\vec{f}_\Lambda ,m) \cdot y }
\left \vert {\partial f_\Lambda \over \partial f } \right \vert^{1/2}   
D^j_{\nu \mu} [R_{wl} (\Lambda ,f)].
\end{equation}
Taking matrix elements give the $ISL(2,C)$ ${\cal D}$-function
\begin{equation}
{\cal D}^{mj}_{f'f}[\Lambda ,Y]  =
e^{i p(\vec{f}' ,m) \cdot y } 
D^j_{\mu' \mu} [ R_{wl} (\Lambda ,f)]
\left \vert {\partial f_\Lambda \over \partial f } \right \vert^{1/2}   
\delta^3 (f' - f_\Lambda ). 
\end{equation}

The infinitesimal generators of $ISL(2,C)$ in this representation can be 
computed using (\ref{eq:JF}-\ref{eq:JH}).  The results are:
\begin{equation}
\hat{P}^{\mu} = p^{\mu}(\vec{f},m)
\end{equation}
\begin{equation}
\hat{J}^j =  i \epsilon^{jkl} {\partial f^m \over \partial p^k } {\partial
\over \partial f^m} \hat{p}^l + (\hat{c}_1^{jk}(p) + i \epsilon_{jlm} 
\hat{c}_2^{lk}(p) \hat{p}^m) 
\hat{j}^k 
\end{equation}
\begin{equation}
\hat{K}^j =  -{1 \over 2} {\partial f^m \over \partial \hat{p}^k }
[ \Delta f^m, \hat{H}]_+
+
i (\hat{c}_1^{jk}(p) - H \hat{c}_2^{jk}(p)) \hat{j}^k 
\end{equation}
where
\begin{equation}
\hat{c}_1^{jk}(p) = {1 \over 2} \mbox{Tr} (L^{-1}(\hat{p}) \sigma_j L (\hat{p}) 
\sigma_k )
\end{equation}
\begin{equation}
\hat{c}_2^{jk}(p) = \mbox{Tr} (L^{-1}(\hat{p}) {\partial \over \partial p^j } 
L (\hat{p}) \sigma_k ).
\end{equation}

These equations can be inverted to obtain explicit expressions 
(\ref{eq:JJA}) for 
$\Delta \hat{f}^k$ in terms of the generators
\begin{equation}
\Delta \hat{f}^k = -{i \over 2\hat{H}} {\partial \hat{H} \over \partial f^k} 
-{1 \over \hat{H}}  \left [ {\partial p^j \over \partial f^k} (\hat{K}^j
- i (\hat{c}_1^{jm}(p) - \hat{H} \hat{c}_2^{jm}(p)) \hat{j}^m  ) \right ] 
\end{equation}
for $k=1,2$ or $3$.  This expression reduces \cite{polyzou} to 
the Newton-Wigner 
position operator when $f^i=p^i$ and the $l$-spin is the canonical spin.
The $l$-spin is given as a function of the
infinitesimal generators by (\ref{eq:APAA}). The partial derivatives 
in this expression are computed with functions which are replaced 
by the appropriate operators after the differentiation is performed.

The $\Delta f^4$ for the 
spins are the raising and lowering operators 
\begin{equation}
\hat{j}_{l \pm} := \hat{j}_{lx} \pm i \hat{j}_{ly} .
\end{equation}
 
This shows explicitly the equivalence between 
\begin{equation}
\lbrace \hat{H} , \vec{\hat{P}}, \vec{\hat{J}}, \vec{\hat{K}} \rbrace
\qquad \mbox{and} \qquad
\lbrace \hat{M}, \hat{j}^2, \vec{\hat{F}}, \Delta \vec{\hat{F}} \rbrace . 
\end{equation}
The basis choices illustrated above, while restrictive, include all
of the basis choices that lead to Dirac's forms of dynamics.  The general 
construction yields a Dirac instant form of dynamics if $f^i$ are taken as 
the three components of the linear momentum and $L_l(p)$ is a canonical 
(rotationless) boost. Dirac's point-form dynamics is obtained if $f^i$ 
is taken as the three components of the four velocity and $L_l(p)$ is the 
canonical boost.  A front form dynamics is obtained if $f^i$ is taken as
the three generators of translations tangent to a light front and $L_l$ is
taken as corresponding the light front boost.  Infinitely many other choices
of $f^i$ and $L_l(p)$ are possible.

\section{Appendix II} 

The Clebsch-Gordan coefficients for the representations in Appendix 1
can be computed from the tensor product representation using the same
methods that were used to construct the single irreducible
representations.  The first step is to decompose the tensor product
representation of the ``rest state'' into irreducible representation
of $SU(2)$.  This requires generalized Melosh rotations to ensure that
all of the spins undergo the same rotations.  The irreducible
representation are then boosted with the appropriate $l$-boost.  This
generally leads to Wigner rotations.  The general result is derived in
\cite{keister}.  The resulting Clebsch-Gordan coefficients for this
basis are:
\[
\langle \vec{f}_1, \mu_1, \vec{f}_2, \mu_2 
\vert \vec{f}, \mu ;m, j , l, s  \rangle =
\]
\[
\delta (\vec{f} -\vec{f} (\vec{f}_1,\vec{f}_2) ) 
\delta (m -m (\vec{f}_1,\vec{f}_2,m_1,m_2) )
\left \vert {\partial (f,k) \over \partial (f_1, f_2)  } \right \vert^{1/2}
{1 \over k} {\partial k \over \partial m} \times
\]
\[   
D^{j_1}_{\mu_1 \mu_1'}  [
R_{wl} (p,k_1) R_{ml} (k_1)]
D^{j_2}_{\mu_2 \mu_2'}  [ 
R_{wl} (p,k_2) R_{ml} (k_2)]
\times
\]
\[
Y^l (\hat{k}_1 (f_1 ,f_2) ) \langle j_1,\mu_1,j_2,\mu_2 
\vert s, \mu_s \rangle \langle s,\mu_s,l,\mu_l \vert j,\mu \rangle 
\]
where $L_c (p)$ is the canonical boost and $L_l(p)$ is a $l$-boost, 
\begin{equation}
k_i = {1 \over 2} \mbox{Tr} 
(L_l^{-1} (p) p_i^{\mu}  \cdot \sigma_{\mu} (L_l^{-1}(p))^{\dagger} )    
\end{equation}
and 
\begin{equation}
R_{wl} (p,k_i):= L_l^{-1}(p_i) L_l(p) L_l(k_i)
\end{equation}
\begin{equation}
R_{mlc} (k_i) := L_l^{-1} (k_i) L_c (k_i).
\end{equation}
These are the Wigner and Melosh rotations associated with the 
$l$-boost.

The Racah coefficient for this choice of basis can be computed in terms of 
four Clebsch-Gordan coefficients.   It is simplest to compute the invariant 
part of this coefficient by choosing $p=(m,0,0,0)$ and integrating the result 
over $SU(2)$.   The Racah coefficients for the couplings
$((12)(3)) \to ((23)(1))$ become:
\[
\langle \vec{f}\,', \mu' ;m',j', (12,3) \vert \vec{f}, 
\mu ;m,j, (23,1) \rangle =
\]
\[
\delta_{j'j} \delta_{\mu' \mu} \delta (\vec{f}'-\vec{f} )
\delta (m-m') 
{1 \over 2j+1} \sum_{\mu_f}  \times 
\]
\[
\left [ {
8 \pi^2 m_{12}m_{23} \omega_2 (q_3'+ q_1)
\over
k_1' k_2 q_3' q_1 \omega_1 (k_1') \omega_2 (k_1') \omega_2( k_2)
\omega_3 (k_2) \omega_{12} (q_3 ) \omega_{23} (q_1)
} \right ] \times
\]
\[
\left [ 
\vert {\partial (f_{12}', f_3') \over \partial (f', q_3')}\vert
\vert {\partial (f_{12}', k_1') \over \partial (f_1', f_2')}\vert
\vert {\partial (f_{23}, k_2) \over \partial (f_2, f_3)}\vert
\vert {\partial (f_{23}, f_1) \over \partial (f, q_1)}\vert
\right ]^{1/2} \times
\]
\[
\langle j , \mu_f \vert L', \mu_L', S', \mu_S' \rangle
\langle S' , \mu_S' \vert j_{12}', \mu_{12}', j_3', \mu_3' \rangle \times
\]
\[
D^{j_{12}}_{\mu_{12}' \mu_{12}}  [ 
R_{mcl} (-q_3') ] Y^{L'*}_{\mu_L'}(\hat{q}_3')
\langle j'_{12} , \mu_{12} \vert l', \mu_l', s', \mu_s' \rangle
\langle s' , \mu_s' \vert j_1', \mu_1', j_2', \mu_2' \rangle 
Y^{l'*}_{\mu_l'}(\hat{k}_1') \times
\]
\[
D^{j_3}_{\mu_3' \mu_3}  [
R_{mcl} (q_3') R_{wl} (-q_1,k_3) R_{mlc} (k_3)]
D^{j_1}_{\mu_1 \mu_1'}  [
R_{mcl} (k_1') R^{-1}_{wl} (-q_3',q_1') R_{ml} (q_1) ]
\times
\]
\[
D^{j_2}_{\mu_2 \mu_2'}  [ 
R_{mcl} (k_2') R^{-1}_{wl} (-q_3',q_2') R_{wl} (-q_1, k_2) R_{mlc} (k_2) ]
\]
\[
Y^{l*}_{\mu_l'}(\hat{k}_2)
\langle j_2 \mu_2 j_3 \mu_3 \vert s \mu_s \rangle
\langle l, \mu_l, s ,\mu_s \vert j_{23} \mu_{23} \rangle 
Y^{L}_{\mu_L}(\hat{q}_1) \times
\]
\[
D^{j_{23}}_{\mu_{23} \mu_{23}'}  [R_{ml} (-q_1)]
\langle j_{23} \mu_{23} j_1 \mu_1 \vert S \mu_S \rangle
\langle L, \mu_L, S ,\mu_S \vert j \mu_f \rangle
\]
where $m$ is the three body invariant mass, $m_{ij}$ are the invariant 
masses of the $ij$ and $jk$ subsystems, $w(k)$ are energies, and $q_i$ 
are the operators
\begin{equation}
\hat{q}_i := L^{-1}_l(\hat{p})\hat{p}_i .
\end{equation}

\section {Appendix III}

To prove Theorem 1 first note that condition (\ref{eq:FAG}) implies 
\begin{equation}
\lim_{t \to \pm \infty} \Vert \hat{U}[I,-T]\left ( \hat{\Phi}_{\cal A} - 
\hat{U}^{\dagger} [\Lambda ,Y] 
\hat{\Phi}_{\cal A} \hat{U}_{\cal A}
[\Lambda ,Y] \right ) \hat{U}_{\cal A} [I,T] \vert \psi \rangle \Vert 
=0
\end{equation}
which is equivalent to 
\begin{equation}
\Omega_{\pm} (\hat{H}, \hat{\Phi}_{\cal A} , \hat{H}_{\cal A} ) = 
\Omega_{\pm} (\hat{H}, \hat{U}^{\dagger} [\Lambda ,Y] 
\hat{\Phi}_{\cal A} \hat{U}_{\cal A} [\Lambda ,Y] , \hat{H}_{\cal A} ). 
\label{eq:APCA}
\end{equation}
Since the Hamiltonian commutes with the linear and angular momentum 
operators, it follows that if $(\Lambda , A)$ is a rotation or translation
this becomes
\begin{equation}
\Omega_{\pm} (\hat{H}, \hat{\Phi}_{\cal A} , \hat{H}_{\cal A} ) = 
\hat{U}^{\dagger}[R,0] 
\Omega_{\pm} (\hat{H}, \hat{\Phi}_{\cal A} , \hat{H}_{\cal A} )
\hat{U}_{\cal A} [R,0]
\label{eq:APCD}
\end{equation}
and 
\begin{equation}
\Omega_{\pm} (\hat{H}, \hat{\Phi}_{\cal A} , \hat{H}_{\cal A} ) = 
\hat{U}^{\dagger}[I,Y] 
\Omega_{\pm} (\hat{H}, \hat{\Phi}_{\cal A} , \hat{H}_{\cal A} )
\hat{U}_{\cal A}[I,Y].
\label{eq:APCE}
\end{equation}

For the case of a rotationless Lorentz transformation condition
(\ref{eq:APCA}) implies
\begin{equation}
\Omega_{\pm} (\hat{H}, \hat{\Phi}_{\cal A} , \hat{H}_{\cal A} )= 
\Omega_{\pm} (\hat{H}, \hat{U}^{\dagger}[\Lambda,0] \hat{\Phi}_{\cal A} 
\hat{U}_{\cal A} [\Lambda,0] , \hat{H}_{\cal A} ).
\end{equation}
The commutation relations imply
\begin{equation}
\hat{U}^{\dagger} [\Lambda ,0] \hat{H}  \hat{U} [\Lambda ,0] = \Lambda^0{}_{\mu} \hat{P}^{\mu} 
\end{equation}
\begin{equation}
\hat{U}_{\cal A}^{\dagger} [\Lambda ,0] \hat{H}_{\cal A}  
\hat{U}_{\cal A} [\Lambda ,0] = \Lambda^0{}_{\mu} \hat{P}_{\cal A}^{\mu}. 
\end{equation}
It follows that 
\[
\hat{U}^{\dagger} [\Lambda ,0]\Omega_{\pm} (\hat{H}, \hat{\Phi}_{\cal A} , 
\hat{H}_{\cal A} )
\hat{U}_{\cal A}[\Lambda ,0]=
\]
\[
\hat{U}^{\dagger} [\Lambda ,0]
\Omega_{\pm} (\hat{H}, \hat{U}[\Lambda,0] \hat{\Phi}_{\cal A} 
\hat{U}_{\cal A}^{\dagger} [\Lambda,0] , \hat{H}_{\cal A} )\hat{U}_{\cal A} 
[\Lambda ,0]=
\]
\begin{equation}
\Omega_{\pm} (\Lambda^0_\mu \hat{P}^{\mu} , \hat{\Phi} , 
\Lambda^0_\mu \hat{P}_{\cal A}^{\mu} )
\label{eq:APCB}
\end{equation}
which can be expressed as
\begin{equation}
\Omega_{\pm} (\Lambda^0_\mu \hat{P}^{\mu} , 
\hat{\Phi}_{\cal A} , \Lambda^0_\mu 
\hat{P}_{\cal A}^{\mu} )=
\lim_{t \to \pm \infty} e^{i \hat{H} \Lambda^0{}_0 t+ i \Lambda^0_{i}
\hat{P}_i t}
\hat{\Phi}_{\cal A}  e^{-i \hat{H}_{\cal A}  \Lambda^0{}_0 t+ i \Lambda^0_{i}\hat{P}_{{\cal A}i} t}.
\label{eq:APCBB}
\end{equation}
Since $\Lambda^0{}_0> 0$ it is possible to redefine define 
$t \to t'= \Lambda^0{}_0t$ so
the limit $t \to \pm \infty$ is equivalent to the 
limit the $t'\to \pm \infty$.  This gives  
\begin{equation}
\lim_{t' \to \pm \infty} e^{i \hat{H} t'}  \hat{U}[I,At'] \hat{\Phi}_{\cal A} 
\hat{U}_{\cal A} [I,At'] 
e^{-i \hat{H}_{\cal A} t'} 
\end{equation}
where 
\begin{equation}
A= {\Lambda^0_i \over \Lambda^0{}_0} \sigma_i .
\end{equation}
Condition (\ref{eq:FAH}) then gives

\begin{equation}
\lim_{t' \to \pm \infty} e^{i \hat{H} t'}  \hat{U}[I,At] \hat{\Phi}_{\cal A} 
\hat{U}_{\cal A} [I,At'] 
e^{-i \hat{H}_{\cal A} t'} =
\Omega_{\pm} (\hat{H}, \hat{\Phi} , \hat{H}_{\cal A} ).
\label{eq:APCC}
\end{equation}
Combining (\ref{eq:APCB}) and (\ref{eq:APCC}) gives 
\begin{equation}
\Omega_{\pm} (\hat{H}, \hat{\Phi}_{\cal A} , \hat{H}_{\cal A} ) = 
\hat{U}^{\dagger} [\Lambda,0] 
\Omega_{\pm} (\hat{H}, \hat{\Phi}_{\cal A} , \hat{H}_{\cal A} )
\hat{U}_{\cal A} [\Lambda,0] .
\label{eq:APCF}
\end{equation}

To complete the proof of Theorem 1 note that 
(\ref{eq:APCD}),(\ref{eq:APCF}), (\ref{eq:APCF})
imply 
\begin{equation}
\hat{U} [\Lambda,Y] \Omega_{\pm} (\hat{H}, \hat{\Phi}_{\cal A} , 
\hat{H}_{\cal A} ) = 
\Omega_{\pm} (\hat{H}, \hat{\Phi}_{\cal A} , \hat{H}_{\cal A} )
\hat{U}_{\cal A} [\Lambda,Y]
\label{eq:APCG}
\end{equation}
which is the intertwining relation of corollary 1.  Corollary 2 follows by 
identifying (\ref{eq:APCBB}) and (\ref{eq:APCC}).

It follows that
\[
\hat{U}^{\dagger}_{\cal A} [\Lambda ,Y] \hat{S} 
\hat{U}_{\cal A} [\Lambda ,Y] =
\]
\[
\hat{U}^{\dagger}_{\cal A} [\Lambda ,Y] 
\Omega^{\dagger}_{+} (\hat{H}, \hat{\Phi}_{\cal A} , 
\hat{H}_{\cal A} )
\Omega_{-} (\hat{H}, \hat{\Phi}_{\cal A} , 
\hat{H}_{\cal A} )
\hat{U}_{\cal A} [\Lambda ,Y] =
\]
\begin{equation}
\Omega^{\dagger}_{+} (\hat{H}, \hat{\Phi}_{\cal A} , 
\hat{H}_{\cal A} )
\hat{U}^{\dagger} [\Lambda ,Y]
\hat{U} [\Lambda ,Y]
\Omega_{-} (\hat{H}, \hat{\Phi}_{\cal A} , 
\hat{H}_{\cal A} )
= \hat{S}.
\end{equation}
This completes the proof of Theorem 1.

To prove corollary 3 note
that equation (\ref{eq:FAJ}) is equivalent to 
\begin{equation}
s-\lim_{s \to \pm \infty} [e^{-i \hat{M} s}\Omega_{\pm} 
(\hat{H}, \hat{\Phi}, \hat{H}_{\cal A}) - \hat{\Phi} 
e^{-i \hat{M}_{\cal A} s}] =0
\end{equation}
The intertwining properties that follow from 
Theorem 1 give 
the strong limit: 
\begin{equation}
s-\lim_{s \to \pm \infty} [(\Omega_{\pm} (\hat{H}, \hat{\Phi}, 
\hat{H}_{\cal A}) - \hat{\Phi}) ]
e^{-i \hat{M}_{\cal A} s} =0.
\label{eq:APCXX}
\end{equation}
The proof that this holds on the dense set of asymptotic states
with bounded momentum follows the proof of theorem IX.23 of \cite{simon2}
(see also \cite{kato}\cite{cg}).  The extension to the strong limit 
follows the argument in \cite{fcwp}.

\section{Appendix IV}

To prove Theorem 4 let $\hat{C}_{a_i}$ be the scattering
equivalence that maps $\hat{U}_{a_i}[\Lambda, Y]$ to the representation
$\tilde{U}_{a_i}[\Lambda , Y]$ with kinematic $\hat{F}^i_{a_j}$, $\Delta
\hat{F}^i_{a_j}$, $\hat{j}_{a_j}$.  Define
\begin{equation}
\hat{C}_a := \otimes_{i=1}^{n_a} \hat{C}_{a_i}
\label{eq:KE}
\end{equation}
and 
\begin{equation}
\tilde{U}_{a}[\Lambda, Y]:= 
\hat{C}_a \hat{U}_{a}[\Lambda, Y] \hat{C}_a^{\dagger} =
\otimes_{i=1}^{n_a} (\hat{C}_{a_i} \hat{U}_{a_i}[\Lambda, Y] 
\hat{C}_{a_i}^{\dagger}).
\label{eq:KF}
\end{equation}

By assumption, the representations $\hat{U}_a [\Lambda ,A]$ and $\tilde{U}_a
[\Lambda ,A]$ have the same scattering matrix elements, which are
products of the single cluster scattering matrix elements.  In
addition, because
\begin{equation}
(\hat{I}-\hat{C}_a) \hat{U}_0 [I,T] = 
\otimes_{i=1}^{n_a} (I_{a_i} - \hat{C}_{a_i})\hat{U}_{0a_i} [I,T]
\label{eq:KG}
\end{equation}
it follows that 
\begin{equation}
\lim_{t \to \pm \infty} (\hat{I}-\hat{C}_a) \hat{U}_0 [I,T] = 0
\label{eq:KH}
\end{equation}
which shows that $\hat{C}_a$ is a scattering equivalence 
on the N-body Hilbert
space.

The representation $\tilde{U}_a [\Lambda ,Y]$ does not have kinematic
$\hat{F}_{i}$, $\Delta \hat{F}_{i}$ or $\hat{j}$, even though
each factor of the tensor product has this property.  The advantage of 
the representation $\tilde{U}_a[\Lambda ,Y]$ is that it is scattering 
equivalent to a representation $\bar{U}_a [\Lambda ,Y]$ that  
has a kinematic $\hat{F}^{i}$, $\Delta \hat{F}^{i}$ and $\hat{j}$.

To show this consider the structure of the single cluster  
$\tilde{H}_{a_i}$ and $\tilde{M}_{a_i}$.  The Hamiltonian  $\tilde{H}_a$
of the representation $\tilde{U}_a [\Lambda ,Y]$ is
\begin{equation}
\tilde{H}_a := \sum_{i=1}^{n_a} \tilde{H}_{a_i}\otimes \hat{I}_{i}
\label{eq:KI}
\end{equation}
where $\hat{I}_i$ is the identity on the remaining factors in the tensor
product.  The mass operator $\tilde{M}_a$ is a function of the
commuting operators $\tilde{M}_{a_i}\otimes \hat{I}_{i}$ and
$\tilde{\vec{P}}_{a_i}\otimes \hat{I}_{i}$.  Corollary 3 of 
Theorem 1 give mild conditions on the interactions for
$\tilde{H}_a$ and $\tilde{M}_a$ to lead to the same $S$-matrix.

The matrix elements of  
$\tilde{M}_{a_i}\otimes \hat{I}_i$ in the tensor product of $n_a$ 
free particle irreducible representations have the form
\[
\langle \otimes_j (f_j, d_j ; m_j, j_j ) \vert
\tilde{M}_{a_i}\otimes \hat{I}_i \vert \otimes_k (f_k', d_k' ;m_k', j_k' )
\rangle =
\]
\[
\delta [f_i, f_i'] \delta_{j_i j_i'} 
\langle d_i ,m_i \Vert
\tilde{M}^{j_i}_{a_i}\Vert d_i', m_i'
\rangle \times 
\]
\begin{equation}
\prod_{k \not= i}
\delta [f_k, f_k'] \delta [d_k, d_k']\delta_{j_k j_k'} 
\delta (m_k -m_k') .
\label{eq:KJ}
\end{equation}

An irreducible free particle basis for the $N$-body system 
can be constructed by successive use
of the $ISL(2,C)$ Clebsch-Gordan coefficients to decompose 
the basis $\vert
\otimes_j (f_j, d_j (m_j, j_j )) \rangle$ into a direct integral of
irreducible representations.  What is relevant for the proof of this
theorem is that the variables $m_i,d_i$ and $j_i$ that appear in the
kernel $\langle d_i ,m_i \Vert
\tilde{M}^{j_i}_{a_i}\Vert d_i', m_i'
\rangle$
of $\tilde{M}_{a_j}$ are degeneracy parameters in this representation.

In order to be precise assume that the irreducible free particle
basis is obtained by successively coupling clusters in the order $(\cdots
(((12)3)4) \cdots n_a)$.  In addition, at each stage in the coupling 
define $\hat{q}_{i}$ as the solution to 
\begin{equation}
\hat{M}_{0(\cdots (12))3) \cdots i+1)}=
\sqrt{\hat{q}_{i}^2 +\hat{M}_{0(i+1)}^2} + 
\sqrt{\hat{q}_{i}^2 + \hat{M}^2_{0(\cdots (12))3) \cdots i)}} .
\label{eq:KK}
\end{equation}
The operators $\hat{q}_i$ are alternate labels for 
the kinematic invariant masses 
$\hat{M}_{0(\cdots (12))3) \cdots i)}$.

Define the single cluster mass operators $\bar{M}_{a_i}$ 
in this irreducible representation
\[
\langle f, d ;m, j \vert \bar{M}_{a_i} \vert
f', d' ;m', j' \rangle =
\label{eq:KL}
\]
\[
\delta [f,f']\delta[j,j']
\delta_{j_i j_i'} 
\langle d_i ,m_i \Vert
\tilde{M}^{j_i}_{a_i}\Vert d_i', m_i'
\rangle \times
\]
\begin{equation}
JJ'
\prod_{k \not= i}
\delta_{j_k j_k'} 
\delta (m_k -m_k')  
\prod_{l=1}^{n_a} (q_l - q_l') \delta_{r_l r_l'} 
\label{eq:KM}
\end{equation}
where the $q_l$'s are considered functions of the kinematic invariant 
masses, the $r_l$ are degeneracy parameters that result when particle 
$l$ is coupled to the irreducible $(1 \cdots l-1)$ system,  
and $J$ and $J'$ are Jacobians 
\begin{equation}
J= \vert {\partial (q_1 \cdots q_{n_a-1} ) \over 
\partial (m_{(12)}\cdots m_{0(\cdots ((12) \cdots (n_a)}}) \vert^{1/2} .
\label{eq:KN}
\end{equation}

The three important observations about this definition are

\begin{itemize} 

\item The non-trivial part of this kernel is identical to the 
non-trivial part of the kernel of $\tilde{M}_{a_i}$ in the tensor product
representation (\ref{eq:KJ}).

\item Each $\bar{M}_{a_j}$ commutes with $\hat{F}_{0}^i$,
$\Delta \hat{F}_{0}^{i}$, $\hat{j}_{0}$

\item $[\bar{M}_{a_i}, \bar{M}_{a_j} ]=0$ 

\end{itemize} 

The relations (\ref{eq:KK}) can be inverted to express the free mass as
a function of the free single cluster mass operators and the $q_i$'s:
\begin{equation}
\hat{M}_0 =  M (\hat{M}_{01}, \cdots ,\hat{M}_{0n_a}, \hat{q}_1, 
\cdots , \hat{q}_{n_a-1})
\label{eq:KP}
\end{equation}

The commutation relations allow the definition: 
\begin{equation}
\bar{M}_a:=  M (\tilde{M}_{a_1}, \cdots ,\tilde{M}_{a_{n_a}}, 
\hat{q}_1, \cdots , \hat{q}_{n_a-1})
\label{eq:KQ}
\end{equation}
where the $\hat{q}_i$'s in (\ref{eq:KQ}) are identical to the
non-interacting $\hat{q}_i$'s in
(\ref{eq:KP}).
By construction $\bar{M}_a$ commutes with  $\hat{F}_{0}^{i}$,
$\Delta \hat{F}_{0}^{i}$, $\hat{j}_{0}$.  Simultaneous 
eigenstates of $\bar{M}_a$, $\hat{F}_0^i$ and $j_0$ transform as
mass $\bar{M}_a$ spin $j_0$ irreducible representations of $ISL(2,)$.
This defines the representation $\bar{U}_a [\Lambda ,Y]$.

In order to construct a scattering theory we need to define a suitable
injection operator to the asymptotic Hilbert space for $\hat{M}_a$.  
The channel injection 
operator for the representation $\tilde{U}_a [\Lambda ,Y]$  
is the tensor product of irreducible eigenstates
\begin{equation}
\tilde{\Phi}_{\alpha} = \vert f_1, \alpha_1 , \cdots ,f_{n_a}, \alpha_1 
\rangle .
\label{eq:KR}
\end{equation}
The corresponding channel injection operator for the representation
$\bar{U}_a [\Lambda ,A]$ is defined as the simultaneous eigenstates of
$\bar{M}_a$, $\hat{j}^2_0$, $\hat{F}^j_0$, $\hat{q}_{i0}$,
$\hat{r}_i$, and $\tilde{M}_{a_i}$ corresponding the same bound states
of the $\tilde{M}_{a_i}$:
\begin{equation}
\bar{\Phi}_{\alpha} = \vert f, j_0, q_1, \cdots ,q_{n_a-1},r_1, \cdots , r_{n_a-1} , 
\alpha_1 \cdots \alpha_{n_a} \rangle .
\label{eq:KS}
\end{equation}
These differ by the delta functions that multiply the cluster eigenfunctions.

With this definition it follows that
\begin{equation}
\bar{\Omega}_{a\pm} := \Omega_{\pm} (\bar{M}_a, 
\bar{\Phi}_{{\cal A}a}, H_{{\cal A}a})
\label{eq:KT}
\end{equation}
exist and are complete.  The scattering operator
\begin{equation}
\bar{S}_a =  \bar{\Omega}^{\dagger}_{a+}\bar{\Omega}_{a-}
= \delta_{j_0 j_0'} 
\delta [f, f'] \prod_{i=1}^{n_a-1}  
\delta (q_i -q_i')\delta [r_i, r'_i] \prod_i \delta_{j_i j_i'} 
\hat{S}_{a_i} 
\label{eq:KTT} 
\end{equation}
is identical to $\tilde{S}_a$ if the Clebsch-Gordan coefficients are 
used to replace the irreducible spectator variables by the 
single cluster $f_i, j_i$'s.  The equivalence follows because the
$S$ matrix elements are determined by the single cluster mass operators,
which have identical reduced kernels 
in representations (\ref{eq:KJ}) and (\ref{eq:KM}).

This establishes that the representations $\tilde{U}_a[\Lambda ,Y]$
and $\bar{U}_a[\Lambda ,Y]$ give the same scattering matrix elements.
By Theorem 3 they are scattering equivalent.  Let $\hat{D}_a$ be the
scattering equivalence that relates these two representations:
\begin{equation}
\bar{U}_a[\Lambda ,Y] =
\hat{D}_a \tilde{U}_a[\Lambda ,Y] \hat{D}^{\dagger}_a .
\label{eq:KU}
\end{equation}
It follows from (\ref{eq:KF}) and (\ref{eq:KU}) that 
\begin{equation}
\bar{U}_a[\Lambda ,Y] = \hat{B}_a 
\hat{U}_a[\Lambda ,Y] \hat{B}_a^{\dagger}
\label{eq:KV}
\end{equation}
where 
\begin{equation}
\hat{B}_a := 
\hat{D}_a \hat{C}_a .
\label{eq:KW}
\end{equation}  
The operator $\hat{B}_a$ is a scattering equivalence since it is a product
of scattering equivalences. This completes the proof of Theorem  4.

\vfill\eject

\end{document}